\documentstyle[titlepage,aps,preprint,psfig,dcolumn]{revtex}
\newcommand{\Li}[1]{\mbox{Li}_2\left(#1\right)}
\newcommand{\bea}{\begin{eqnarray}}
\newcommand{\eea}{\end{eqnarray}}
\newcommand{\be}{\begin{equation}}
\newcommand{\ee}{\end{equation}}

\newcommand{\Tr}{{\rm Tr}}

\newcommand{\<}{\langle\,}
\renewcommand{\>}{\,\rangle}

\newcommand{\sslash}{s\!\!\!/}
\newcommand{\tr}{\mbox{Tr}}
\renewcommand{\Re}{\mbox{Re}}
\renewcommand{\Im}{\mbox{Im}}

\newcommand{\bfsig}{{\mbox{\boldmath $\sigma$}}}

\newcommand{\one}{1\!\mbox{l}}
\newcommand{\as}{\alpha_s}
\newcommand{\eq}[1]{(\ref{#1})}
\newcommand{\nn}{\nonumber}
\newcommand{\sumw}{\widehat{\sum}}
\newcommand{\VAp}{V\!\!A_{+}}

\renewcommand{\thefootnote}{\fnsymbol{footnote}}

\makeatletter
\draft

                \def\@preprint{}
                \def\preprint#1#2#3{%
                \ifpreprintsty
                \def\@preprint{
                \noindent \hbox{#1}\hfill\hbox{#2}\\
                \hbox{#3}\vskip -2ex}%
                \fi
                }

\begin{document}

%
%

\preprint{hep-ph/9806306}{PITHA 98/13}{}
\title{The spin density matrix of top quark pairs 
produced in electron-positron
annihilation including QCD radiative corrections\footnote{supported  
by BMBF, contract 057AC9EP.}
}
\author{
        Arnd Brandenburg\footnote{supported  by Deutsche
                                Forschungsgemeinschaft.}
, Marc Flesch
, and Peter Uwer
        }
\address{
        Institut f\"ur Theoretische Physik, RWTH Aachen,
        D-52056 Aachen, Germany
        }
\date{\today}
\maketitle{}
\begin{abstract}
We calculate the spin density matrix of top quark pairs 
for the reaction $e^+e^-\to t\bar{t}X$ to order $\alpha_s$. 
As an application we show  next-to-leading order 
results for a variety
of spin observables for the $t\bar{t}$ system. 
These include the top quark
and antiquark polarizations and $t\bar{t}$ spin-spin correlations 
as a function of the center-of-mass energy and 
of the top quark 
scattering angle for arbitrary longitudinal 
polarization of the electron/positron beam.
\end{abstract}
\bigskip

\bigskip
\newpage
\setcounter{footnote}{0}
 \renewcommand{\thefootnote}{\arabic{footnote}}
\setcounter{page}{1}
%
%
\begin{section}{Introduction}\label{sec1}
Among the six known quark flavours known to date,
the top quark is of particular interest: Its large mass
implies that very high energies are involved in the 
production and decay of this particle, which in turn 
allows for tests of the fundamental interactions at
these high energy scales. Moreover, the interactions
of the top quark can be studied in greater detail than 
those of the lighter particles since the top quark essentially 
behaves like a free, but extremely short-lived particle.
With a mass of $m \approx 175$ GeV, the 
lifetime of the top quark is 
about $5\times 10^{-25}$ seconds. This short lifetime
effectively cuts off the long distance QCD dynamics.
In particular, the top quark polarization is not diluted by
hadronization and thus becomes an additional observable to
test perturbative QCD, or, more generally, short 
distance physics.
\par
An ideal machine to study the properties of top quarks in detail
would be a high-luminosity, high-energetic $e^+e^-$ linear collider.
The physics potential of such a machine is described for example
in \cite{Ac97}. We just mention here that at  center-of-mass energies
in the range $\sqrt{s}= 400-1000$ GeV, an annual yield of the 
order of $10^5$ top quark pairs may be expected.
\par 
For the process $e^+e^-\to t\bar{t}X$,
the production cross sections for longitudinally  \cite{KoPiTu94}
and transversely \cite{BeMaSc92} polarized top quarks
are known to order $\alpha_s$. The correlations between the spins
of top quarks and  antiquarks have been studied 
extensively in leading
order  \cite{PaSh96}. 
The {\it longitudinal} spin-spin correlations
have also been calculated in next-to-leading order (NLO)
\cite{TuBePe98,GrKoLe98}. Polarization phenomena in top quark pair
production near threshold have been investigated in \cite{Ha95}. 
\par 
A convenient theoretical framework to discuss spin phenomena
is the concept of the spin density matrix, and  
the main objective of this paper is to 
present results for the full spin 
density matrix of the $t\bar{t}$ system to order $\alpha_s$.
This allows for a systematic study of spin effects
in $e^+e^-\to t\bar{t}X$.
For phenomenological applications, our results 
should be  supplemented by 
the decay matrices at NLO for the different $t$ and $\bar{t}$ 
decay channels  \cite{CzJeKu91,La98}.  
\par
An alternative approach to the analysis of spin effects in top quark
production and decay is the computation of the relevant 
helicity amplitudes. This was accomplished at next-to-leading order 
in \cite{Sc96}, where also a Monte Carlo event generator for the case
of semileptonic $t\bar{t}$ decays was constructed.
\par
The outline of the rest of this paper is as follows. We start 
in section \ref{sec2} by
introducing the spin density matrix formalism and apply it
to the reaction $e^+e^-\to t\bar{t}$ at leading order. In section \ref{sec3} we compute the  
QCD radiative corrections to the results of section \ref{sec2}.
Section \ref{sec4} contains numerical 
results for a variety of spin observables.
We exhibit their dependence on the c.m. energy 
and on the top quark scattering angle and
further study the effects of electron beam polarization.
\end{section}
\begin{section}{Kinematics and leading order results}\label{sec2}
\setcounter{equation}{0}
  In this section we review some basic kinematics and the concept of the
  spin density matrix formalism. To set up the notation, we start with a 
  closer look at the amplitude for the process
  \begin{equation}
    \label{process} e^+(p_+)e^-(p_-)\to (\gamma^\ast,Z^\ast)\to t(k_t)
    \bar t(k_{\bar t}) X,
  \end{equation}
  where $e^-$($e^+$) denotes an electron (positron) and $t$($\bar t$) 
  describes a  top (anti-) quark with mass $m$. 
We work in
  leading order in the electroweak coupling and in next-to-leading order in
  the strong coupling $\as=g_s^2/(4\pi)$. To this order the unspecified 
  rest $X$ can be only a gluon.
  The amplitude for the reaction (\ref{process}) can be written in the 
  following form:
  \begin{eqnarray}
    \label{amplitude}
    {\cal T}_{fi} &=& \frac{4\pi\alpha}{s}\Big\{\,\chi(s)\, \bar v(p_+)
    (g_v^e  \gamma_{\mu} - g_a^e 
    \gamma_{\mu}\gamma_5) u(p_-)\,
    (g_v^t V^\mu - g_a^t A^\mu ) 
    + \bar v(p_+) \gamma_{\mu} u(p_-) (- Q_t V^{\mu})\Big\}.\nn\\
  \end{eqnarray} 
  In (\ref{amplitude}), $s=(p_++p_-)^2$, $Q_t$ denotes 
  the electric charge of the top quark in units of $e=\sqrt{4\pi\alpha}$, and
  $g_v^f$, $g_a^f$ are the vector- and the axial-vector couplings of a
  fermion of type $f$, i.e.
  \bea
    g_v^f = T_3^f - 2\, Q_f\,\sin^2\vartheta_W,\quad
    \mbox{and} \quad g_a^f= T_3^f,
  \eea
  in particular $g_v^e = -\frac{1}{2} + 2 \sin^2\vartheta_W$, 
  $g_a^e =-\frac{1}{2}$ for an electron, and  
  $g_v^t = \frac{1}{2} - \frac{4}{3} \sin^2\vartheta_W$,
  $g_a^t = \frac{1}{2}$ for a top quark, with $\vartheta_W$ denoting the 
  weak mixing angle. The function
  $\chi(s)$ is given by
  \begin{equation}
    \label{chi}
    \chi(s) = \frac{1}{4\sin^2\vartheta_W\cos^2\vartheta_W}\,
    \frac{s}{s-m_Z^2 + i m_Z \Gamma_Z},
  \end{equation}
  where $m_Z$ and $\Gamma_Z$ stand for the mass and 
  the width of the Z boson.
  (We keep here the width of the Z boson because it will be 
  relevant for an application of
  our results to b quark production at the Z resonance.)
  The amplitudes $V_\mu, A_\mu$ in \eq{amplitude} encode the 
  information on the decay of the vector boson 
into the $t\bar t$ and $t\bar tg$
  final states. In particular they depend on the momentum and the polarization
  of the outgoing particles. 
  Considering only longitudinal polarization for the incoming electrons 
  and/or positrons and neglecting the lepton masses leads to 
  \begin{eqnarray}
    \label{master}
    |{\cal T}_{fi}|^2 &=& \frac{16\pi^2\alpha^2}{s^2} 
    \Big[L^{PC\mu\nu}H^{PC}_{\mu\nu}+L^{PV\mu\nu}H^{PV}_{\mu\nu}\Big]
  \end{eqnarray}
  for the square of \eq{amplitude}.
  The lepton tensors $L^{PC(PV)\mu\nu}$ read
    \begin{eqnarray}
    \label{leptontensors}
    L^{PC\mu\nu}= p_{+}^{\mu}p_{-}^{\nu}+p_{+}^{\nu}p_{-}^{\mu}
    -g^{\mu\nu}p_+p_-,\quad &\mbox{and}&\quad
    L^{PV\mu\nu}= 
    -i\varepsilon^{\mu\nu}_{\ \ \rho\sigma}p_+^{\rho}p_-^{\sigma}.
  \end{eqnarray}
  The tensors $H^{PC(PV)}_{\mu\nu}$ describing the decay of a polarized 
  Z boson can be written as
  \begin{equation}
    \label{master2}
    H^{PC(PV)}_{\mu\nu} \ =\       
    g^{VV}_{PC(PV)} H^{VV}_{\mu\nu} + g^{AA}_{PC(PV)} H_{\mu\nu}^{AA}
    + g_{PC(PV)}^{VA_+} H_{\mu\nu}^{VA_+} 
    + g_{PC(PV)}^{VA_-}H_{\mu\nu}^{VA_-},      
  \end{equation}
  with
  \begin{equation}
    \label{tensors}
    H_{\mu\nu}^{VV} = V_\mu V_\nu^\ast, \ 
    H_{\mu\nu}^{AA} = A_\mu A_\nu^\ast, \ \mbox{and}\ 
    H_{\mu\nu}^{VA_\pm} = V_\mu A_\nu^\ast \pm A_\mu V_\nu^\ast.
  \end{equation}
  The couplings $g_X^Y$ ($X\in\{PC,PV\}$, $Y\in\{VV,AA,VA_+,VA_-\}$)  in (\ref{master2}) are given by
  \begin{eqnarray}
    \label{wcouplings}
    g_{PC (PV)}^{VV} &=&Q_t^2\, f_{PC(PV)}^{\gamma\gamma} 
    + 2 \,g_v^t\,Q_t\, \Re \chi(s)\,f_{PC(PV)}^{\gamma Z} 
    + g_v^{t\,2}\, |\chi(s)|^2\,f_{PC(PV)}^{ZZ},\nonumber\\
    g_{PC(PV)}^{AA}  &=& g_a^{t\,2} |\chi(s)|^2 f_{PC(PV)}^{ZZ},
    \nonumber\\
    g_{PC(PV)}^{VA_+}&=& -g_a^t\,Q_t\,\Re\,\chi(s)\,f_{PC(PV)}^
    {\gamma Z} -g_v^t\,g_a^t\, |\chi(s)|^2 f_{PC(PV)}^{ZZ},\nonumber\\
    g_{PC(PV)}^{VA_-}&=&i\,g_a^t\,Q_t\,\Im\chi(s) f_{PC(PV)}^{\gamma Z},
  \end{eqnarray}
  where
  \begin{equation}
    \begin{array}[h]{lcl}
      \label{fcouplings}
      f_{PC}^{ZZ}&=&(1-\lambda_-\lambda_+)(g_v^{e2}+g_a^{e2})-
      2(\lambda_--\lambda_+) g_v^{e} g_a^{e},\\
      f_{PV}^{ZZ}&=&(\lambda_--\lambda_+) (g_v^{e\,2}+g_a^{e\,2}) -
      2\,(1-\lambda_-\lambda_+)g_v^e\,g_a^e,\\
      f_{PC}^{\gamma Z}&=&-(1-\lambda_-\lambda_+)g_v^e + 
      (\lambda_--\lambda_+)g_a^e,\\
      f_{PV}^{\gamma Z}&=& (1-\lambda_-\lambda_+)g_a^e -(\lambda_--\lambda_+)
      g_v^e,
    \end{array}
    \qquad
    \begin{array}[h]{lcl}
      f_{PC}^{\gamma\gamma}&=&1-\lambda_-\lambda_+,\\
      f_{PV}^{\gamma\gamma}&=& \lambda_--\lambda_+,\\
      &&\\
      &&
    \end{array}
  \end{equation}
  with $\lambda_-$ ($\lambda_+$) denoting the longitudinal 
  polarization  of the electron (positron) beam\footnote{
    For a right-handed electron (positron), $\lambda_{\mp}=+1$.}.
  The couplings $g_{PC(PV)}^{VA_-}$ are formally 
of higher order in the 
  electroweak couplings. 
  The structure $H_{\mu\nu}^{VA_-}$ will therefore 
  not be discussed further. For top quark production, 
  where $\sqrt{s}\gg m_Z$,  
  one should set the width $\Gamma_Z$  of the Z boson to zero for 
  consistency.

  The (unnormalized) spin density matrix for the reaction (\ref{process}) 
  may be defined by
  \begin{eqnarray}
    \label{rho0}
    \rho_{\alpha\alpha',\beta\beta'} &=&  \sumw
    \langle t(k_t,\alpha)\bar t(k_{\bar t},\alpha')X|{\cal T}|
    e^+(p_+,\lambda_+)e^-(p_-,\lambda_-)\rangle\nn\\
    &&\ \ \ \  \langle t(k_t,\beta)\bar t(k_{\bar t},\beta')X|{\cal T}|
    e^+(p_+,\lambda_+)e^-(p_-,\lambda_-)\rangle^\ast,
  \end{eqnarray}
  where $\alpha,\alpha',\beta,\beta'$ are the spin indices of the outgoing 
  top (anti-) quarks. The sum $\sumw$ in \eq{rho0} runs over all unobserved 
  degrees of freedom such as the colour of the outgoing particles or the
  polarization of the emitted gluon.
  In \eq{rho0} one should read the combination 
  $\alpha\alpha'$ ($\beta\beta'$) on the left-hand side as a shorthand
  notation for a multi-index built from $\alpha,\alpha'$ ($\beta,\beta'$).
  To calculate the spin density matrix it is convenient to use a different
  representation which follows immediately from the concept of the
  density matrix:
  \begin{equation}
    \label{rho1}
\sumw\left|{\cal T}(e^+(p_+,\lambda_+)e^-(p_-,\lambda_-)
\to t(k_t,\hat{\bf s}_t)\bar 
    t(k_{\bar t},\hat{\bf s}_{\bar t})X)\right|^2 
    = \tr\left[\rho\cdot \frac{1}{2}(\one+\hat{\bf s}_{t}\cdot\bfsig)
      \otimes\frac{1}{2}(\one+\hat{\bf s}_{\bar t}\cdot\bfsig)
    \right].
  \end{equation}
  Here $\hat{\bf s}_t\ (\hat{\bf s}_{\bar t})$ is the unit
polarization  of the 
  top
  (anti-) quark in the rest frame of the top (anti-) quark\footnote{We define 
the rest frame of the outgoing top (anti-) quark as 
the rest system which is obtained by a rotation-free Lorentz-boost 
    from the center-of-mass system of the $e^+e^-$-pair.}, and 
  $\sigma_i$ are the usual Pauli matrices. With $\otimes$ we denote
  the tensor product between the spin space of the quark and the antiquark.
  Using in \eq{rho1} a decomposition of the spin density matrix $\rho$ of the 
  form
  \begin{equation}
    \label{rho2}
    \rho = a\ \one\otimes \one 
    + {\bf B}^+ \cdot \bfsig\otimes\one
    + \one\otimes\bfsig\cdot {\bf B}^- 
    + C_{ij}\sigma_i\otimes\sigma_j,
  \end{equation}
  the density matrix can be easily calculated by a comparison
  of the polarization independent parts, terms proportional to $\hat {\rm s}_{ti}$ 
  ($\hat{\rm s}_{\bar ti}$),
  and terms proportional to $\hat{\rm s}_{ti} \hat{\rm s}_{\bar tj}$ 
  on the left-hand side and the right-hand side of \eq{rho1}.   
  More precisely we define
  \begin{equation}
    \label{rho3}
    \rho = 4\pi^2\alpha^2N_C\,  \sum_{Y,X} g_X^Y \rho_Y^X
  \end{equation}
  ($X\in\{PC,PV\},\ Y\in\{VV,AA,VA_+\}$), with
  \begin{equation}
    \tr\left[\rho_Y^X\cdot \frac{1}{2}(\one+\hat{\bf s}_{t}\cdot\bfsig)
      \otimes\frac{1}{2}(\one+\hat{\bf s}_{\bar t}\cdot\bfsig)
    \right] = \frac{1}{N_C}\frac{4}{s^2}\sumw L^{X \mu\nu} H_{\mu\nu}^{Y},
  \end{equation}
  where $N_C$ is the number of colours, and $g_X^Y$ are the couplings as 
  given in \eq{wcouplings}.
  For the density matrices $\rho_Y^X$ we use a representation as in 
  \eq{rho2}. It is useful to decompose the polarizations ${\bf B}_Y^{X,\pm}$ 
  and the spin-spin correlations $C_{Y,ij}^X$ further. 
  For the two-parton final state it is convenient to write:
  \begin{eqnarray}
    \label{rho_comp}
    {\bf B}^{\pm} &=& b_1^{\pm} \hat{\bf p} + 
b_2^{\pm} \hat{\bf k} + b_3^{\pm} \hat{\bf n},
    \nonumber\\
    C_{ij} &=& c_0 \delta_{ij} +\varepsilon_{ijk}(c_1 \hat{\mbox{p}}_k + 
    c_2\hat{\mbox{k}}_k +c_3\hat{\mbox{n}}_k)
    +c_4\hat{\mbox{p}}_i\hat{\mbox{p}}_j + c_5 \hat{\mbox{k}}_i\hat{\mbox{k}}_
    j+
    c_6(\hat{\mbox{p}}_i\hat{\mbox{k}}_j+\hat{\mbox{p}}_j\hat{\mbox{k}}_i)
    \nonumber\\
    &+&c_7(\hat{\mbox{p}}_i\hat{\mbox{n}}_j+\hat{\mbox{p}}_j\hat{\mbox{n}}_i) 
    + c_8(\hat{\mbox{k}}_i\hat{\mbox{n}}_j+\hat{\mbox{k}}_j\hat{\mbox{n}}_i),
  \end{eqnarray}
  with 
  \begin{equation}
      \hat{\bf p} = \frac{\displaystyle{\bf p}_- }
      {\displaystyle|{\bf p}_-|},\ \ 
      \hat{\bf k} = \frac{\displaystyle{\bf k}_t}
      {\displaystyle|{\bf k}_t|},\ \ 
      \hat{\bf n} = \frac{\displaystyle\hat{\bf p}\times\hat{\bf k}} 
      {\displaystyle|\hat{\bf p}\times\hat{\bf k}|},
  \end{equation}
where the three-momenta  ${\bf p}$ and ${\bf k}$ are defined in
$e^+e^-$ c.m. system. 
  In \eq{rho_comp} we suppress for simplicity the additional indices $Y,X$. 
  For the case of the three-parton final state a similar decomposition
  can be used. 
  A detailed discussion of the properties of $\rho$ under discrete
  symmetry transformations is given in \cite{BeNaOvSc92}.
  In leading order ($O(\as^0)$) the non-vanishing entries in the 
  density matrices $\rho^X_Y$ read:
  \begin{equation}
    \begin{array}[h]{lclcl}  
      \label{rhoborn}
      a^{PC}_{VV}=2-\beta^2(1-z^2) ,&&
      a^{PC}_{AA}=\beta^2(1+z^2) ,&&
      b^{\pm, PC}_{1,V\!\!A_{+}}=2\beta rz ,\\
      c_{0,VV}^{PC}=-\beta^2(1-z^2) ,&&
      c_{0,AA}^{PC}=\beta^2(1-z^2) ,&&
      b^{\pm,PC}_{2,V\!\!A_{+}}=2\beta\Big(1+(1-r)z^2\Big) ,\\
      c_{4,VV}^{PC}=2 ,&&
      c_{4,AA}^{PC}=-2\beta^2 ,&&\\
      c_{5,VV}^{PC}=2\Big((1-r)^2z^2+\beta^2\Big) ,&&
      c_{6,AA}^{PC}=2\beta^2z ,&&\\
      c_{6,VV}^{PC}=-2(1-r)z,&&&&\\
      \\
      b^{\pm,PV}_{1,VV}=2r ,&&
      b^{\pm,PV}_{2,AA}=2\beta^2z ,&&
      a^{PV}_{V\!\!A_{+}}=4\beta z ,\\
      b^{\pm,PV}_{2,VV}=2(1-r)z ,&&&&
      c_{5,V\!\!A_{+}}^{PV}=4\beta(1-r)z ,\\
      &&&&c_{6,V\!\!A_{+}}^{PV}=2\beta r,
    \end{array}
  \end{equation}
  where $z=\hat {\bf p}\cdot\hat {\bf k}$, 
  $\beta = \sqrt{1-4m^2/s}$, and $r= 2m/\sqrt{s}$.

  The leading order differential 
  cross section $d\sigma_0(\hat{\bf s}_t,\hat{\bf s}_{\bar t})$
  is related to the leading order density 
  matrix $\rho_0$ as follows:
  \begin{equation}
    \label{dsigma}
    d\sigma(\hat{\bf s}_t,\hat{\bf s}_{\bar t}) = \frac{1}{2s}
    \tr\left[\rho_0
\cdot \frac{1}{2}(\one+\hat{\bf s}_{t}\cdot\bfsig)
      \otimes\frac{1}{2}(\one+\hat{\bf s}_{\bar t}\cdot\bfsig)
    \right]dR_2
  \end{equation}
  with
  \begin{equation}
    \label{r2}
    dR_2 = \frac{d^3k_t}{(2\pi)^3 2k_{t}^0}
    \frac{d^3k_{\bar t}}{(2\pi)^3 2k_{\bar t}^0} 
    (2\pi)^4\delta(p_++p_--k_t-k_{\bar t})
  \end{equation}
  The total cross section for example can be obtained from 
  \begin{equation}
  \sigma_0=\frac{1}{2s}\frac{\beta }{16\pi}\int\limits_{-1}^{1} \!dz\,\tr[\rho_0]
  = \frac{1}{2s}\pi\alpha^2 N_C\beta\int\limits_{-1}^{1} \!dz\,
  (g_{PC}^{VV}\, a_{VV}^{PC} + g_{PC}^{AA}\, a_{AA}^{PC}),    
  \end{equation}
 yielding the well known result:
  \begin{equation}
    \label{eq:sigmapt}
    \sigma_0 = \sigma_{pt} N_C \beta
        \left(\frac{3-\beta^2}{2}g_{PC}^{VV}+\beta^2 g_{PC}^{AA}\right),
        \quad \mbox{with}\quad
        \sigma_{pt} =
    \frac{4\pi\alpha^2}{3s}.
  \end{equation}
  Within the framework of the spin density matrix formalism it is easy to
  calculate spin observables. For instance, at leading order 
  the polarization of the top quark projected onto its momentum direction
  can be obtained from:
  \begin{eqnarray}
    \label{example1}
    \langle \hat {\bf k}\cdot{\bf S}_t\rangle &=&
    \frac{\int\limits_{-1}^{1}\!dz\,
      \tr\left[\rho_0\cdot\left(\hat {\bf k}\cdot\frac{\bfsig}{2}\otimes \one\right)\right]}
    {\int\limits_{-1}^{1}\!dz\,\tr[\rho_0]}\nn\\
    &=&\frac{2\int\limits_{-1}^{1}\!dz\,
      g_{PC}^{\VAp}(z b_{1,\VAp}^{+,PC} + b_{2,\VAp}^{+,PC})
      +g_{PV}^{VV}(z b_{1,VV}^{+,PV} + b_{2,VV}^{+,PV})
      +g_{PV}^{AA} b_{2,AA}^{+,PV}}
    {4\int\limits_{-1}^{1}\!dz\,
      (g_{PC}^{VV}\, a_{VV}^{PC} + g_{PC}^{AA}\, a_{AA}^{PC})}\nn\\
    &=&\frac{2\beta g_{PC}^{\VAp} }
   { (3-\beta^2)g_{PC}^{VV}+2\beta^2 g_{PC}^{AA}},
  \end{eqnarray}
where ${\bf S}_t= \frac{\bfsig}{2}\otimes \one $ is the top quark spin operator.
(The spin operator of the top antiquark is ${\bf S}_{\bar{t}}= \one\otimes \frac{\bfsig}{2}$.)
  As another example  
consider the following spin-spin correlation,   
  which is in leading order proportional to the so-called 
longitudinal spin-spin 
correlation studied in \cite{TuBePe98,GrKoLe98}:
  \begin{eqnarray}
    \label{example2}
    \langle
    (\hat {\bf k}\cdot{\bf S}_t)(\hat {\bf k}\cdot{\bf S}_{\bar t})\rangle 
    &=&\frac{\int\limits_{-1}^{1} \!dz\,
      \tr\left[\rho_0\cdot \left(\hat {\bf k}\cdot\frac{\bfsig}{2}
      \otimes \hat {\bf k}\cdot\frac{\bfsig}{2}\right)\right]}
    {\int\limits_{-1}^{1}\!dz\,\tr[\rho_0]}
    =\frac{1}{4}\frac{(1+\beta^2)g_{PC}^{VV} 
      + 2\beta^2 g_{PC}^{AA}}{(3-\beta^2)g_{PC}^{VV}+2\beta^2 g_{PC}^{AA}}.
  \end{eqnarray}

 The examples above show  that the spin density matrix formalism 
 enables one to calculate efficiently 
the expectation values of spin observables. 
A more exhaustive analysis of spin observables 
together with next-to-leading order numerical results will be 
presented in section \ref{sec4}. 
\end{section}
\begin{section}{QCD radiative corrections}\label{sec3}
\setcounter{equation}{0}
The QCD corrections 
at order $\alpha_s$ 
to the expectation values of spin observables
are given by the contributions
from one-loop virtual 
corrections to $e^+e^-\to t\bar{t}$ and 
from the real gluon emission 
process $e^+e^-\to t\bar{t}g$ at leading order. 
We first give some details on the computation 
of the virtual corrections. 

Both infrared (IR) and ultraviolet (UV) singularities which
appear in the one-loop integrals of the virtual corrections
are treated within the framework of dimensional regularization
in $d=4-2\epsilon$ space-time dimensions.
We use the 't Hooft-Veltman prescription \cite{HoVe79} to treat the 
$\gamma_5$ matrix present in the axial vector current part
of the vertex correction in $d$ dimensions. 
It is well known that this prescription violates 
certain Ward identities.
They are restored by adding 
a finite counterterm \cite{La93}.
The UV singularities are removed by appropiate 
counterterms fixed by on-shell renormalization conditions 
for the quark.
After renormalization one obtains UV finite 
vertex corrections for the vector and the axial vector 
parts of the amplitude to order $\alpha_s$.
\par
The renormalized amplitude still contains an IR       
singularity which appears as a single pole in $\epsilon$
and which multiplies -- up to a factor -- the Born amplitude.
This singularity is cancelled in infrared safe quantities 
by a corresponding singularity from the real gluon emission
process. The latter singularity 
is obtained from the phase space integration
of the squared matrix element for $e^+e^-\to t\bar{t}g$
over the region of phase space where the gluon is soft. 
\par
The virtual corrections  to the density
matrix are obtained by first computing the interference
between the renormalized one-loop amplitude and the
Born amplitude for given polarization vectors $\hat{\bf s}_{t},\ 
\hat{\bf s}_{\bar{t}}$
and then extracting  $\rho^{\rm virtual}$
as described in section \ref{sec2} below equation (\ref{rho1}).
Note that the necessary trace algebra can now be 
performed in $d=4$ dimensions without punity. 
In particular, the projectors 
$(1+\gamma_5\sslash_{t,\bar{t}})/2$ can be
kept in 4 dimensions. 
\par
We now discuss the contributions from real gluon emission.
We isolate the soft gluon singularities by splitting
the $t\bar{t}g$ phase space into a soft and a hard gluon region. 
The soft gluon region is defined by the condition 
\bea
E_g\le x_{\rm min}\frac{\sqrt{s}}{2},
\eea
where $E_g$ is the gluon energy
in the c.m. system and $x_{\rm min}$ is a sufficiently small
quantity.
The hard gluon region is the complement 
of the soft region.
In the limit where the gluon momentum $k_g$ goes to zero
one can neglect $k_g$ in the numerator of 
${\cal T}_{fi}(e^+e^-\to t\bar{t}g)$, which leads to 
\begin{eqnarray}\label{soft1}
\rho(e^+e^-\to t\bar{t}g) {\buildrel
k_g\to 0\over \longrightarrow}
4\pi\alpha_sC_F\left\{ \frac{2k_tk_{\bar t}}{(k_tk_g)(k_{\bar{t}}k_g)}
-\frac{m^2}{(k_tk_g)^2}-\frac{m^2}{(k_{\bar t}k_g)^2}  \right\}
\rho_0(e^+e^-\to t\bar{t}).
\end{eqnarray}
Using (\ref{soft1})
in the whole soft gluon region   
leads to the approximation 
\begin{eqnarray}\label{soft}
\int\!\frac{d^{d-1}k_g}{(2\pi)^{d-1}2E_g}
\Theta(x_{\rm min}\frac{\sqrt{s}}{2}-E_g)
\rho(e^+e^-\to t\bar{t}g) \approx S \rho_0(e^+e^-\to t\bar{t})
\equiv\rho^{\rm soft},
\end{eqnarray}
where the soft factor $S$ is given by
\bea
\label{softfac}
S&=& 4\pi\alpha_sC_F\int\!\frac{d^{d-1}k_g}{(2\pi)^{d-1}2E_g}
\Theta(x_{\rm min}\frac{\sqrt{s}}{2}-E_g)
\left\{ \frac{2k_tk_{\bar t}}{(k_tk_g)(k_{\bar{t}}k_g)}
-\frac{m^2}{(k_tk_g)^2}-\frac{m^2}{(k_{\bar t}k_g)^2}  \right\}
\nn \\ &=& 
\frac{\alpha_s}{2\pi}C_F\frac{1}{\Gamma(1-\epsilon)}
\left(\frac{4\pi\mu^2}{s}\right)^{\epsilon}
\left(x_{\rm min}^2\right)^{-\epsilon}
\frac{1}{\epsilon}\frac{1}{\beta}
\Bigg\{ 2\beta+(1+\beta^2)\ln(\omega) \nonumber \\
&-&2\epsilon
\Bigg[
\ln(\omega)
+(1+\beta^2)\left(
\Li{1-\omega}+\frac{1}{4}\ln^2(\omega)
\right)
\Bigg]\Bigg\} +{\cal O}(\epsilon).
\eea

Here, $C_F=(N_C^2-1)/(2N_C)$, $\beta=\sqrt{1-4m^2/s}$, and  
$\omega = (1-\beta)/(1+\beta)$.
The scale $\mu$ is introduced in (\ref{softfac})
to keep the strong coupling constant dimensionless in $d$ dimensions.
The dependence 
on $\mu$ cancels in the sum of the virtual and soft contributions.
 
For finite $x_{\rm min}$, the sum of the contributions from the
soft and hard gluon region differs from the exact result by
terms of order $x_{\rm min}$ because of the soft gluon approximation. 
The sum becomes exact for $x_{\rm min}\to 0$.
With the choice $x_{\rm min}=10^{-5}$ the systematic
error due to this approximation is smaller than one 
permill in all our numerical results. 
This can be nicely checked
by varying $x_{\rm min}$ between, say, $10^{-3}$ and 
$10^{-6}$ and numerically extrapolating to zero. 
\par
The sum of the virtual and soft contributions
to the density matrix $\rho$ is finite and 
can be written in a compact form as follows: 

We  define:
\bea
L&=& -\frac{\alpha_s}{2\pi}C_F\frac{1}{\beta}\Bigg\{\left(
2\beta+(1+\beta^2)\ln(\omega)\right)
\left[\ln\!\left(x_{\rm min}^2\right)
-\ln\left(\frac{1-\beta^2}{4}\right)+2\right]\nn\\
&+& (1+\beta^2)\left(4\Li{1-\omega}
+\ln^2(\omega)-\pi^2\right)\Bigg\},
\eea
and use as further abbreviations
\bea
\kappa = \frac{\alpha_s}{2\pi}C_F,\qquad
\ell_{1}= -\kappa\beta\ln(\omega),\qquad 
\ell_{2}= (2-\beta^2)\ell_1,\qquad
\ell_{3}= \frac{1}{\beta}\ell_1. 
\eea

Then,

\bea 
\lim_{\epsilon\to 0}
\left(\rho^{\rm virtual}+\rho^{\rm soft}\right)
= L\rho_0+ \rho^{\rm rest},
\eea

where the nonvanishing building blocks 
of  $\rho_0$ are listed in equation (\ref{rhoborn})
of section II.
The matrix $\rho^{\rm rest}$ is also decomposed according
to equation (\ref{rho3}) with matrices $\rho_Y^{X,\rm rest}$ 
expanded like in (\ref{rho2}), (\ref{rho_comp}).
The nonvanishing entries of the various matrices 
$\rho_Y^{X,\rm rest}$
that make up  $\rho^{\rm rest}$ read (we suppress here
the index ``rest'' for aesthetic reasons): 
 \begin{equation}
   \begin{array}[h]{lclcl}  
     \label{rhovirt}
a^{PC}_{VV}=(1+z^2)\ell_{1} , && 
a^{PC}_{AA}=(1+z^2)\ell_{2} ,&& 
b^{\pm,PC}_{1,V\!\!A_{+}}=-zr(\beta^2-2)\ell_{3} ,\\
b^{\pm,PC}_{3,VV}=-\kappa\pi r \beta z\sqrt{1-z^2} , &&
c_{0,AA}^{PC}=(1-z^2)\ell_{2} , &&
b^{\pm,PC}_{2,V\!\!A_{+}}=\Big(2(1+z^2)+r(\beta^2-2)z^2\Big)\ell_{3} ,\\
c_{0,VV}^{PC}=-(1-z^2)\ell_{1} ,&&
c_{4,AA}^{PC}=-2\ell_{2} ,&&
c_{7,V\!\!A_{+}}^{PC}=-2\kappa\pi(1-\beta^2)\sqrt{1-z^2} ,\\
c_{4,VV}^{PC}=2\ell_{1} ,&&
c_{6,AA}^{PC}=2z\ell_{2} ,&&
c_{8,V\!\!A_{+}}^{PC}=\kappa\pi
z\Big(2(1\!-\!\beta^2)+(\beta^2\!-\!2)r\Big)\sqrt{1-z^2} ,\\
c_{5,VV}^{PC}=2(1+(1-r)z^2)\ell_{1} , && && \\
c_{6,VV}^{PC}=-z(2-r)\ell_{1} , && && \\ \\
b^{\pm,PV}_{1,VV}=r\ell_{1} ,&&
b^{\pm,PV}_{2,AA}=2z\ell_{2} ,&&
a^{PV}_{V\!\!A_{+}}=4z\ell_{3} , \\
b^{\pm,PV}_{2,VV}=z(2-r)\ell_{1} , && &&
b^{\pm,PV}_{3,V\!\!A_{+}}=\kappa\pi r(\beta^2-2)\sqrt{1-z^2} ,\\
c_{8,VV}^{PV}=-\kappa\pi r\beta\sqrt{1-z^2} ,&& &&
c_{5,V\!\!A_{+}}^{PV}=2z\Big(r(\beta^2-2)+2\Big)\ell_{3} ,\\
 &&&& c_{6,V\!\!A_{+}}^{PV}=-r(\beta^2-2)\ell_{3} .\\
 \end{array}
\end{equation}
\par

For a given  observable, the contributions from gluons 
with energy $E_g > x_{\rm min}\sqrt{s}/2$  
are calculated by a 
numerical integration over the hard gluon region of the 
three-body phase space.
The spin density matrix $\rho^{\rm hard}(e^+e^-\to t\bar{t}g)$
for the hard gluon emission process  
is obtained by evaluating the left-hand side of equation 
(\ref{rho1}) for $X= g$. 
The individual matrices $\rho^{X,\rm hard}_Y$ 
are rather lengthy and we do not list them in this paper.
We just mention here that 
instead of the expansion (\ref{rho_comp}) of  
${\bf B}^{\pm},\ C_{ij}$  with respect to
$\hat{\bf p},\ \hat{\bf k},$ and $\hat{\bf n}$ that was
used for the two-parton final state, we found it more 
convenient for the three-parton final state in the hard gluon region 
to use as basis vectors ${\bf k}_t/|{\bf k}_t|,\ 
{\bf k}_{\bar t}/|{\bf k}_{\bar t}|$,
and 
$({\bf k}_t\times {\bf k}_{\bar t})/
|{\bf k}_t\times     {\bf k}_{\bar t}|$.  
Note that the matrix $\rho^{\rm hard}(e^+e^-\to t\bar{t}g)$
does not contain any singularities and that the whole
computation can be performed
in $d=4$ dimensions.

\end{section} 
\begin{section}{Numerical results}\label{sec4}
\setcounter{equation}{0}
In this section we present next-to-leading order 
results for expectation values of a variety of spin observables.
For an observable ${\cal O}$ we  use the notation
\bea 
\label{obsexpand}
\langle {\cal O}\rangle &=&  
\langle {\cal O} \rangle_0 + \frac{\alpha_s}{\pi} 
\langle {\cal O} \rangle_1 
+ O\left( \frac{\alpha_s^2}{\pi^2}\right),\nonumber \\
\sigma &=& \sigma_0 + 
\frac{\alpha_s}{\pi} \sigma_1 + 
O\left( \frac{\alpha_s^2}{\pi^2}\right),  
\eea
where $\sigma$ is the total cross section for 
$e^+e^-\to t\bar{t}X$, and 
\bea
\label{obs01}
\langle {\cal O}\rangle_0 &=& \frac{1}{\sigma_0}
\frac{1}{2s}\int\!dR_2\,\Tr\left\{ 
\rho_0\cdot {\cal O}\right\},\nonumber \\
\langle {\cal O} \rangle_1 &=& 
\frac{1}{\sigma_0}\frac{1}{2s}
\Bigg[
\int\!dR_2 \,\Tr\left\{ 
\lim_{\epsilon\to 0}\left(\rho^{\rm soft}
+\rho^{\rm virtual}\right) \cdot {\cal O}\right\}
\nonumber \\ &+&
\int\!dR_3\, \Theta(E_g-x_{\rm min}\sqrt{s}/2)\Tr\left\{ 
\rho^{\rm hard}\cdot {\cal O}\right\}\Bigg]
- \langle {\cal O}\rangle_0 \frac{\sigma_1}{\sigma_0}.
\eea
Here, $dR_2$ is given in (\ref{r2}) and
\bea 
dR_3 &=& \frac{d^3k_t}{(2\pi)^32k^0_t}
\frac{d^3k_{\bar{t}}}{(2\pi)^32k^0_{\bar{t}}}
\frac{d^3k_g}{(2\pi)^32k^0_g}(2\pi)^4
\delta(p_++p_--k_t-k_{\bar{t}}-k_g). 
\eea
We consider the following set of  observables:
\bea
\label{obslist}
{\cal O}_1 &=& \hat{\bf p}\cdot{\bf S}_t, \ \ \ \ \ \ \ \ \ \ \ \  
\bar{{\cal O}}_1 = \hat{\bf p}\cdot{\bf S}_{\bar t}, \nonumber \\
{\cal O}_2 &=& \hat{\bf k}\cdot{\bf S}_t, \ \ \ \ \ \ \ \ \ \ \ \ 
\bar{{\cal O}}_2 = \hat{\bf k}\cdot{\bf S}_{\bar t}, \nonumber \\
{\cal O}_3 &=& \hat{\bf n}\cdot{\bf S}_t, \ \ \ \ \ \ \ \ \ \ \ \ 
\bar{{\cal O}}_3 = \hat{\bf n}\cdot{\bf S}_{\bar t}, \nonumber \\
{\cal O}_4 &=& {\bf S}_t\cdot{\bf S}_{\bar t}, \nonumber \\
{\cal O}_5 &=&\hat{\bf p}\cdot ( {\bf S}_t\times {\bf S}_{\bar t}) , \nonumber \\
{\cal O}_6 &=&\hat{\bf k}\cdot ( {\bf S}_t\times {\bf S}_{\bar t}) , \nonumber \\
{\cal O}_7 &=&\hat{\bf n}\cdot ( {\bf S}_t\times {\bf S}_{\bar t}) , \nonumber \\
{\cal O}_8 &=& (\hat{\bf p}\cdot{\bf S}_t) (\hat{\bf p}\cdot{\bf S}_{\bar t}), \nonumber \\
{\cal O}_9 &=& (\hat{\bf k}\cdot{\bf S}_t) (\hat{\bf k}\cdot{\bf S}_{\bar t}), \nonumber \\
{\cal O}_{10} &=& (\hat{\bf p}\cdot{\bf S}_t) (\hat{\bf k}\cdot{\bf S}_{\bar t})+
(\hat{\bf k}\cdot{\bf S}_t) (\hat{\bf p}\cdot{\bf S}_{\bar t}), \nonumber \\
{\cal O}_{11} &=& (\hat{\bf p}\cdot{\bf S}_t) (\hat{\bf n}\cdot{\bf S}_{\bar t})+
(\hat{\bf n}\cdot{\bf S}_t) (\hat{\bf p}\cdot{\bf S}_{\bar t}), \nonumber \\
{\cal O}_{12} &=& (\hat{\bf k}\cdot{\bf S}_t) (\hat{\bf n}\cdot{\bf S}_{\bar t})+
(\hat{\bf n}\cdot{\bf S}_t) (\hat{\bf k}\cdot{\bf S}_{\bar t}).
\eea
The expectation values $\langle {\cal O}_2\rangle_0$ 
and $\langle {\cal O}_9\rangle_0$ are given in analytic form in
(\ref{example1}) and (\ref{example2}), respectively.
\par
Several constraints are imposed by discrete symmetries 
on the expectation values of the
observables (\ref{obslist}). An unpolarized $e^+e^-$ initial state is an eigenstate
of the combined charge conjugation (C) and parity (P) transformation. CP invariance 
of the interactions considered here then  
implies $\langle {\cal O}_1 \rangle = \langle \bar{\cal O}_1 \rangle$ and 
$\langle {\cal O}_5 \rangle =0$. Further, differences between  $\langle {\cal O}_2 \rangle$
and  $\langle \bar{\cal O}_2 \rangle$ 
as well as  nonzero values for $\langle {\cal O}_{6} \rangle$ and  $\langle {\cal O}_{7} \rangle$
can only be generated by the contributions
from hard gluon emission, 
since ${\bf S}_t {\buildrel{\rm {CP}} \over {\rightarrow}}  {\bf S}_{\bar t}$, 
${\bf k}_t {\buildrel{\rm {CP}} \over {\rightarrow}} -{\bf k}_{\bar{t}}$,
and  since we have ${\bf k}_{\bar{t}}= -{\bf k}_t$ for a final state
consisting solely of a $t\bar{t}$ pair 
(recall that the three-momenta are defined in the $e^+e^-$ c.m. system). 
From invariance under the time reversal operation T it follows that nonzero  
$\langle {\cal O}_3 \rangle$,  $\langle \bar{\cal O}_3 \rangle$,  
$\langle {\cal O}_6 \rangle$,  $\langle {\cal O}_{11} \rangle$, and  
$\langle {\cal O}_{12} \rangle$ can only be generated by absorptive
parts in the scattering amplitude. To order $\alpha_s$ this means that
$\langle {\cal O}_6 \rangle$ is exactly zero due to CP invariance, while 
$\langle {\cal O}_3 \rangle = \langle \bar{\cal O}_3 \rangle$,   
$\langle {\cal O}_{11} \rangle$, and  
$\langle {\cal O}_{12} \rangle$  get nonzero, albeit small, contributions
from the imaginary parts of the one-loop integrals appearing
in the virtual corrections (cf. the functions $b_{3,VV}^{\pm,PC}$, 
$b_{3,VA_{+}}^{\pm,PV}$, and 
$c_{8,VV}^{PV}$, $c^{PC}_{7,8,VA_{+}}$  of 
equation (\ref{rhovirt})).
All the above arguments also hold for the case of  polarized electrons
(and/or positrons), although in that case the initial state has no definite
CP parity. This is because the net effect of a CP transformation
of the initial state is
$\lambda_{\mp} \to - \lambda_{\pm}$ in our formulas, 
and hence the couplings $g^X_Y$ are left unchanged. 
(cf. (\ref{wcouplings}), (\ref{fcouplings})).
\par
In Table I we list our 
results for the expectation values of (\ref{obslist})
in terms of the quantities 
$\langle {\cal O}_i\rangle_  {0,1}$ as defined in
(\ref{obsexpand}) and (\ref{obs01}). 
We choose four different
c.m. energies, namely $\sqrt{s} = 400$, $500$, $800$, and $1000$ GeV.
The positron beam is always assumed to be unpolarized, while
for the electron beam the three cases $\lambda_-=0,\pm 1$ are
considered.  
As numerical input we use $m_Z = 91.187$ GeV, 
an on-shell top quark mass of
$m = 175$ GeV, and $\sin^2\vartheta_W=0.2236$.

The table shows that the top quark and antiquark
are produced highly polarized and also that the
spin-spin correlations are large.
For example, the polarization\footnote{The polarization
is conventionally defined as {\it two times} the expectation
value of the spin operator.} of the top quark
projected onto the beam axis at $\sqrt{s}=500$ GeV
and for $\lambda_-=+1$ amounts to
\bea
2 \langle \hat{\bf p}\cdot{\bf S}_t \rangle =
0.8998-0.278\frac{\alpha_s}{\pi}=0.8910,
\eea
where we set $\alpha_s=0.1$.
As another example, consider the spin-spin correlation 
$\langle {\cal O}_{10} \rangle$ at $\sqrt{s}=1$ TeV, also
for $\lambda_-=+1$:
\bea
 \langle (\hat{\bf p}\cdot{\bf S}_t)
(\hat{\bf k}\cdot{\bf S}_{\bar t})+
(\hat{\bf k}\cdot{\bf S}_t)
(\hat{\bf p}\cdot{\bf S}_{\bar t})  \rangle =
0.2770-0.303\frac{\alpha_s}{\pi}=0.2674,
\eea 
where we again set $\alpha_s=0.1$.
                   
A global characteristic of all the expectation values of the 
observables (\ref{obslist}) is that the QCD corrections 
are quite small.
The quantity $\alpha_s/\pi\times|\langle {\cal O}_i\rangle_1/
\langle {\cal O}_i\rangle_0|$  ranges, for nonzero 
$\langle {\cal O}_i\rangle_0$ and (a fixed value of) 
$\alpha_s=0.1$ between
$1.9$ permill (for $\langle {\cal O}_4 \rangle$ at 
$\sqrt{s}=400$ GeV and all three choices of $\lambda_-$) 
and $5.3$ percent (amusingly also 
for $\langle {\cal O}_4 \rangle$, but at $\sqrt{s}=1000$ GeV and
$\lambda_-=-1$)\footnote{In leading order, 
$\langle {\cal O}_4 \rangle=1/4$, since the reaction proceeds
through a single spin-one boson.}. 
\par
To check our calculation, we compared our numerical value for
the order $\alpha_s$ correction to the total cross section $\sigma_1$
with the value one gets by using the analytic formula as 
given for example
in \cite{ChKuKw96} and  found excellent agreement.  
Note that the longitudinal spin-spin correlation $\langle P^{\ell\ell}\rangle$
studied in \cite{GrKoLe98} is, at next-to-leading order, {\it not}
proportional to our expectation value 
$\langle {\cal O}_9 \rangle$: 
The former
would correspond in our 
notation to the expectation value
${\protect 4\langle (\hat{\bf k}_t\cdot {\bf S}_t)
(\hat{\bf k}_{\bar t}\cdot {\bf S}_{\bar t}) \rangle}$,
which only at 
leading order is equal to $-4\langle {\cal O}_9 \rangle$.
To compare our results for $\langle P^{\ell\ell}\rangle$,  
we reproduced Figures 1 and 2 of reference 
\cite{GrKoLe98}
and found agreement.
\par
We now study the distributions of our expectation values 
with respect to $z$, the cosine 
of the top quark scattering angle in the c.m. system. 
These distributions
are defined as  
$ \langle{\cal O}_i\delta(z-z') \rangle$,
i.e. we do not average over $z$ but 
over all other kinematic variables.

The distributions 
$\langle{\cal O}_{3,11,12}\delta(z-z') \rangle$ are not shown,
since they can be easily constructed from the listed analytic
formulas for $b^{\pm}_3,c_{7,8}$ of equation (\ref{rhovirt}). 
We also do not
show the distribution  
$\langle{\cal O}_{7}\delta(z-z') \rangle$, since 
according to Table I the expectation
value  $\langle{\cal O}_{7} \rangle$ 
varies (again for a fixed $\alpha_s=0.1$)
between the tiny values $-0.6\times 10^{-4}$ 
and $-0.3$\%. 
 
Figs. 1a and 1b  show, to NLO accuracy,   
the distribution $\langle{\cal O}_{1}\delta(z-z') \rangle$ 
at c.m. energies $\sqrt{s}=500$ GeV and $\sqrt{s}=1$ TeV
, respectively, for $\lambda_-=0,\pm 1$.
In this and all the following plots we set $\alpha_s=0.1$.
Note that the distribution gets more peaked near 
$z =+1$
as the c.m. energy rises. This feature is less
pronounced in the distribution
 $\langle{\cal O}_{2}\delta(z-z') \rangle$ depicted in Figs. 
2a,b.
In Figs. 3 - 6 we show the distributions 
for different spin-spin correlations, namely  
$\langle{\cal O}_{4,8,9,10}\delta(z-z') \rangle$.
In these figures, the c.m. energy is varied
between $\sqrt{s}=400$ GeV and $\sqrt{s}=1$ TeV, while
the electron polarization is set
to $\lambda_-=0$. The results for other choices of 
 $\lambda_-$  do not differ much from the ones shown.
This is also reflected in the rather weak dependence of the
spin-spin correlations $\langle{\cal O}_{4,8,9,10}\rangle$ 
on  $\lambda_-$ (cf. Table I).
Note that the distributions typically rise as $z\to +1$.  
\par
To illustrate the impact of the $O(\alpha_s)$ corrections,
we plot in Figs. 7 - 10 the ``$K$-factors''
\bea \label{kfactor}
K_i(z)= \frac{\langle{\cal O}_i\delta(z-z') \rangle_0
+\alpha_s/\pi
\langle{\cal O}_i\delta(z-z') \rangle_1}
{\langle{\cal O}_i\delta(z-z') \rangle_0}
\eea
for $i=1$ (Figs. 7a,b), and $i=4,8,9$ (Figs. 8,9,10).
The $K$-factors 
show a strong dependence both on the cosine of the 
scattering angle and on the c.m. energy. They vary between
$0.88$ and $1.04$. 
\end{section}
\begin{section}{Conclusions}\label{sec5}
The production of top quark pairs in $e^+e^-$ annihilation
involves a variety of spin phenomena. 
We have performed a systematic study
of these effects to order $\alpha_s$ and including
beam polarization effects 
using the spin density matrix formalism. 
Apart from a
significant polarization of the top quarks and antiquarks,
the spins of $t$ and $\bar{t}$ are also strongly
correlated. 
The QCD corrections to the leading order results
for the expectation values of all spin observables
considered are at the percent level or smaller.
The spin effects in the $t\bar{t}$ production
will manifest themselves in the angular
distributions of the $t$ and $\bar{t}$ decay
products. For a  phenomenological
analysis of these angular distributions, 
one can combine the results presented
in this paper with spin decay matrices 
computed to next-to-leading order accuracy
for the different $t$ and $\bar{t}$ decay modes.

\end{section}
\section*{Acknowledgments}
We would like to thank W. Bernreuther for many
enlightening discussions and for his comments on
the manuscript.

\newpage
\begin{center}
TABLE CAPTION
\end{center}
\noindent
{\bf Table I.} Expectation values of the observables 
listed in (\ref{obslist})
in terms of the quantities 
$\langle {\cal O}_i\rangle_  {0,1}$ as defined in
(\ref{obsexpand}) and (\ref{obs01}) for different c.m. energies, 
$\lambda_+=0$ and $\lambda_-=0,\pm 1$. 
For the expectation values not listed in the table we have,
as discussed in the text,
$\langle \bar{\cal O}_{1,3}\rangle =\langle {\cal O}_{1,3}\rangle$,
and $\langle {\cal O}_{5,6}\rangle=0$.
\newpage
\begin{center}
FIGURE CAPTIONS
\end{center}
\noindent 
{\bf Fig. 1.} Expectation 
value $\<{\cal O}_1 \delta(z-z')\>$ to order $\alpha_s$
for a fixed value $\alpha_s=0.1$ and $\lambda_+=0$. 
In 1a (1b) the c.m. energy is set to $\protect\sqrt{s}=500$ GeV 
($\protect\sqrt{s}=1$ TeV). 
The solid line is the result for $\lambda_-=0$, 
the dashed line for 
$\lambda_-=-1$, and the dotted line for $\lambda_-=+1$.\\
\noindent 
{\bf Fig. 2.} Same as  Fig. 1, but for 
$\<{\cal O}_2 \delta(z-z')\>$. \\
\noindent 
{\bf Fig. 3.} Expectation value $\<{\cal O}_4 \delta(z-z')\>$ 
to order $\alpha_s$ for a fixed value $\alpha_s=0.1$, 
$\lambda_+=\lambda_-=0$, and c.m. energies  
$\protect\sqrt{s}=400$ GeV (dashed line),  
$\protect\sqrt{s}=500$ GeV (solid line),
 $\protect\sqrt{s}=800$ GeV (dotted line), and
 $\protect\sqrt{s}=1000$ GeV (dash-dotted line).\\
\noindent 
{\bf Fig. 4.} Same as  Fig. 3, but for 
$\<{\cal O}_8 \delta(z-z')\>$. \\
\noindent 
{\bf Fig. 5.} Same as  Fig. 3, but for 
$\<{\cal O}_9 \delta(z-z')\>$. \\
\noindent 
{\bf Fig. 6.} Same as  Fig. 3, but for 
$\<{\cal O}_{10} \delta(z-z')\>$. \\
\noindent 
{\bf Fig. 7.} The function $K_1(z)$ 
defined in equation (\ref{kfactor}).  
In 7a (7b) the c.m. energy is set to $\protect\sqrt{s}=500$ GeV 
($\protect\sqrt{s}=1$ TeV). 
The solid line is the result for $\lambda_-=0$, 
the dashed line for 
$\lambda_-=-1$, and the dotted line for $\lambda_-=+1$.\\
\noindent
{\bf Fig. 8.} The function $K_4(z)$ 
defined in equation (\ref{kfactor}) for
$\lambda_+=\lambda_-=0$ and c.m. energies  
$\protect\sqrt{s}=400$ GeV (dashed line),  
$\protect\sqrt{s}=500$ GeV (solid line),
 $\protect\sqrt{s}=800$ GeV (dotted line), and
 $\protect\sqrt{s}=1000$ GeV (dash-dotted line).\\ 
\noindent 
{\bf Fig. 9.} Same as  Fig. 8, but for $K_8(z)$.  \\
\noindent 
{\bf Fig. 10.} Same as  Fig. 8, but for $K_9(z)$.  
\begin{center}
  \begin{table}[htb]
    \ifpreprintsty\renewcommand{\arraystretch}{0.59}\fi
    \begin{tabular}[h]{lc|D{.}{.}{4}|D{.}{.}{4}|D{.}{.}{4}|D{.}{.}{4}
        |D{.}{.}{4}|D{.}{.}{4}|D{.}{.}{4}|D{.}{.}{4}|}
      \hline\hline
      \multicolumn{2}{c|}{}&
      \multicolumn{8}{c|}{}\\[0.6ex]
      \multicolumn{2}{c|}{}&
      \multicolumn{8}{c|}{c.m. energy in GeV}\\[2ex]
      \multicolumn{2}{c|}{}&
      \multicolumn{2}{c|}{400}&
      \multicolumn{2}{c|}{500}&
      \multicolumn{2}{c|}{800}&
      \multicolumn{2}{c|}{1000}\\[1.5ex]
      &
      \multicolumn{1}{c|}{$\lambda_-$}&
      \multicolumn{1}{c|}{$\langle {\cal O}_i\rangle_0$}&
      \multicolumn{1}{c|}{$\langle {\cal O}_i\rangle_1$}&
      \multicolumn{1}{c|}{$\langle {\cal O}_i\rangle_0$}&
      \multicolumn{1}{c|}{$\langle {\cal O}_i\rangle_1$}&
      \multicolumn{1}{c|}{$\langle {\cal O}_i\rangle_0$}&
      \multicolumn{1}{c|}{$\langle {\cal O}_i\rangle_1$}&
      \multicolumn{1}{c|}{$\langle {\cal O}_i\rangle_0$}&
      \multicolumn{1}{c|}{$\langle {\cal O}_i\rangle_1$}\\[1.3ex]
      \hline
&$-$& -0.4870&   0.039& -0.4608&   0.125& -0.4014&   0.309& -0.3760&   0.377 \\
$\langle{\cal O}_{1}\rangle$&0& -0.2048&   0.024& -0.1867&   0.064& -0.1554&   0.133& -0.1438&   0.156 \\
&+&  0.4811&  -0.052&  0.4499&  -0.139&  0.3895&  -0.302&  0.3654&  -0.363 \\
 \hline
&$-$& -0.1686&  -0.191& -0.2578&  -0.216& -0.3397&  -0.118& -0.3581&  -0.059 \\
$\langle{\cal O}_{2}\rangle$&0& -0.0583&  -0.065& -0.0870&  -0.070& -0.1120&  -0.036& -0.1173&  -0.017 \\
&+&  0.2099&   0.225&  0.3094&   0.228&  0.3927&   0.102&  0.4104&   0.040 \\
 \hline
&$-$& -0.1686&  -0.185& -0.2578&  -0.165& -0.3397&   0.113& -0.3581&   0.267 \\
$\langle\bar{\cal O}_{2}\rangle$&0& -0.0583&  -0.063& -0.0870&  -0.053& -0.1120&   0.040& -0.1173&   0.090 \\
&+&  0.2099&   0.218&  0.3094&   0.167&  0.3927&  -0.165&  0.4104&  -0.334 \\
 \hline
&$-$&0
&  -0.332&0
&  -0.232&0
&  -0.121&0
&  -0.093 \\
$\langle{\cal O}_{3}\rangle$&0&0
&  -0.356&0
&  -0.246&0
&  -0.127&0
&  -0.097 \\
&+&0
&  -0.413&0
&  -0.279&0
&  -0.140&0
&  -0.106 \\
 \hline
&$-$&    0.25&  -0.015&    0.25&  -0.087&    0.25&  -0.310&    0.25&  -0.418 \\
$\langle{\cal O}_{4}\rangle$&0&    0.25&  -0.015&    0.25&  -0.086&    0.25&  -0.307&    0.25&  -0.414 \\
&+&    0.25&  -0.015&    0.25&  -0.084&    0.25&  -0.300&    0.25&  -0.405 \\
 \hline
&$-$&0
&  -0.002&0
&  -0.016&0
&  -0.062&0
&  -0.082 \\
$\langle{\cal O}_{7}\rangle$&0&0
&  -0.002&0
&  -0.017&0
&  -0.065&0
&  -0.086 \\
&+&0
&  -0.003&0
&  -0.019&0
&  -0.071&0
&  -0.094 \\
 \hline
&$-$&  0.2392&  -0.033&  0.2240&  -0.095&  0.2006&  -0.224&  0.1927&  -0.280 \\
$\langle{\cal O}_{8}\rangle$&0&  0.2375&  -0.037&  0.2205&  -0.100&  0.1957&  -0.225&  0.1876&  -0.278 \\
&+&  0.2332&  -0.046&  0.2123&  -0.111&  0.1848&  -0.225&  0.1765&  -0.272 \\
 \hline
&$-$&  0.1173&   0.039&  0.1606&   0.030&  0.2128&  -0.142&  0.2258&  -0.244 \\
$\langle{\cal O}_{9}\rangle$&0&  0.1182&   0.041&  0.1620&   0.031&  0.2137&  -0.143&  0.2265&  -0.246 \\
&+&  0.1206&   0.045&  0.1652&   0.032&  0.2157&  -0.147&  0.2279&  -0.249 \\
 \hline
&$-$&  0.1580&   0.162&  0.2191&   0.107&  0.2442&  -0.141&  0.2417&  -0.247 \\
$\langle{\cal O}_{10}\rangle$&0&  0.1693&   0.170&  0.2323&   0.106&  0.2560&  -0.155&  0.2528&  -0.264 \\
&+&  0.1968&   0.189&  0.2630&   0.102&  0.2823&  -0.187&  0.2770&  -0.303 \\
 \hline
&$-$&0
&   0.330&0
&   0.222&0
&   0.097&0
&   0.066 \\
$\langle{\cal O}_{11}\rangle$&0&0
&   0.114&0
&   0.075&0
&   0.032&0
&   0.022 \\
&+&0
&  -0.410&0
&  -0.266&0
&  -0.112&0
&  -0.076 \\
 \hline
&$-$&0
&   0.181&0
&   0.225&0
&   0.189&0
&   0.160 \\
$\langle{\cal O}_{12}\rangle$&0&0
&   0.077&0
&   0.093&0
&   0.076&0
&   0.064 \\
&+&0
&  -0.177&0
&  -0.213&0
&  -0.174&0
&  -0.146 \\
 \hline
 \hline
      \label{tab1} \end{tabular}
\ifpreprintsty\renewcommand{\arraystretch}{1}\fi
\caption{}
\end{table}
\begin{figure}[h]
\unitlength1.0cm
\begin{picture}(16.,8.)
\put(-1,-1){\psfig{figure=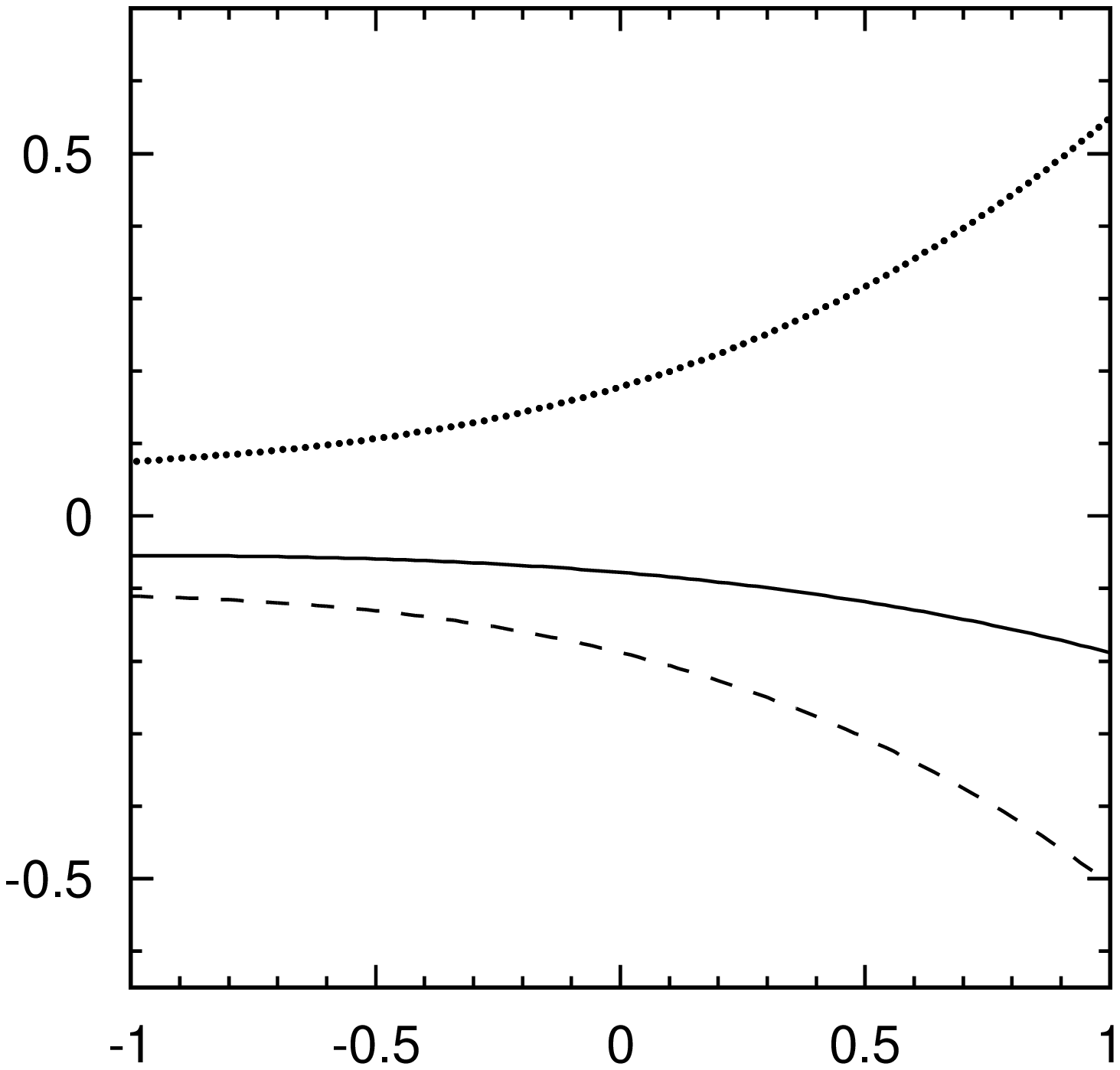,height=9cm,width=9cm}}
\put(8,-1){\psfig{figure=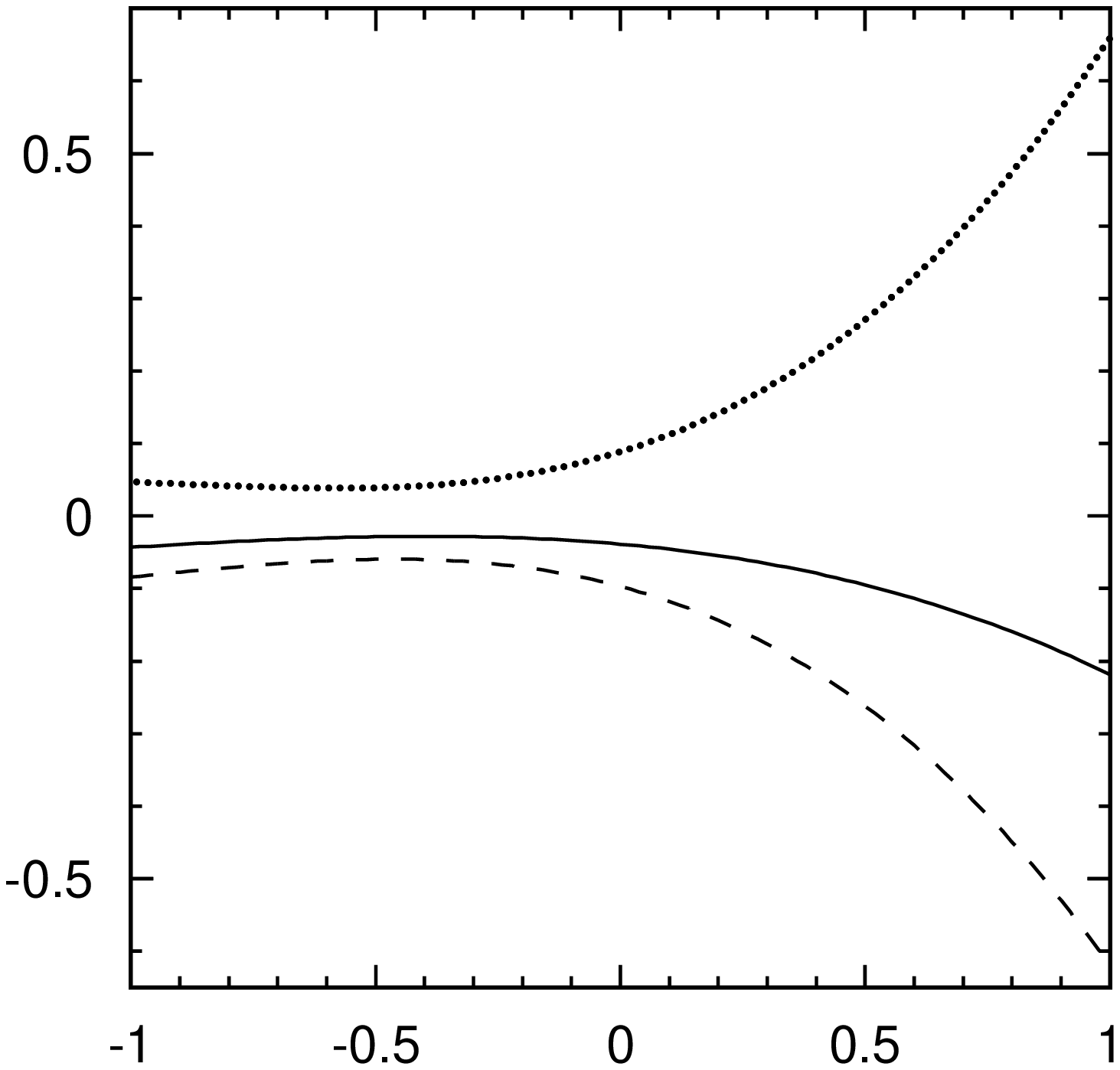,height=9cm,width=9cm}}
\put(0.6,6.2){\mbox{a)}}
\put(9.6,6.2){\mbox{b)}}
\put(3.4,-0.6){\mbox{$z$}}
\put(12.4,-0.6){\mbox{$z$}}
\end{picture}
\vskip 0.5cm
\caption{}\label{fig:o1}
\end{figure}
\begin{figure}[h]
\unitlength1.0cm
\begin{picture}(16.,8.)
\put(-1,-1){\psfig{figure=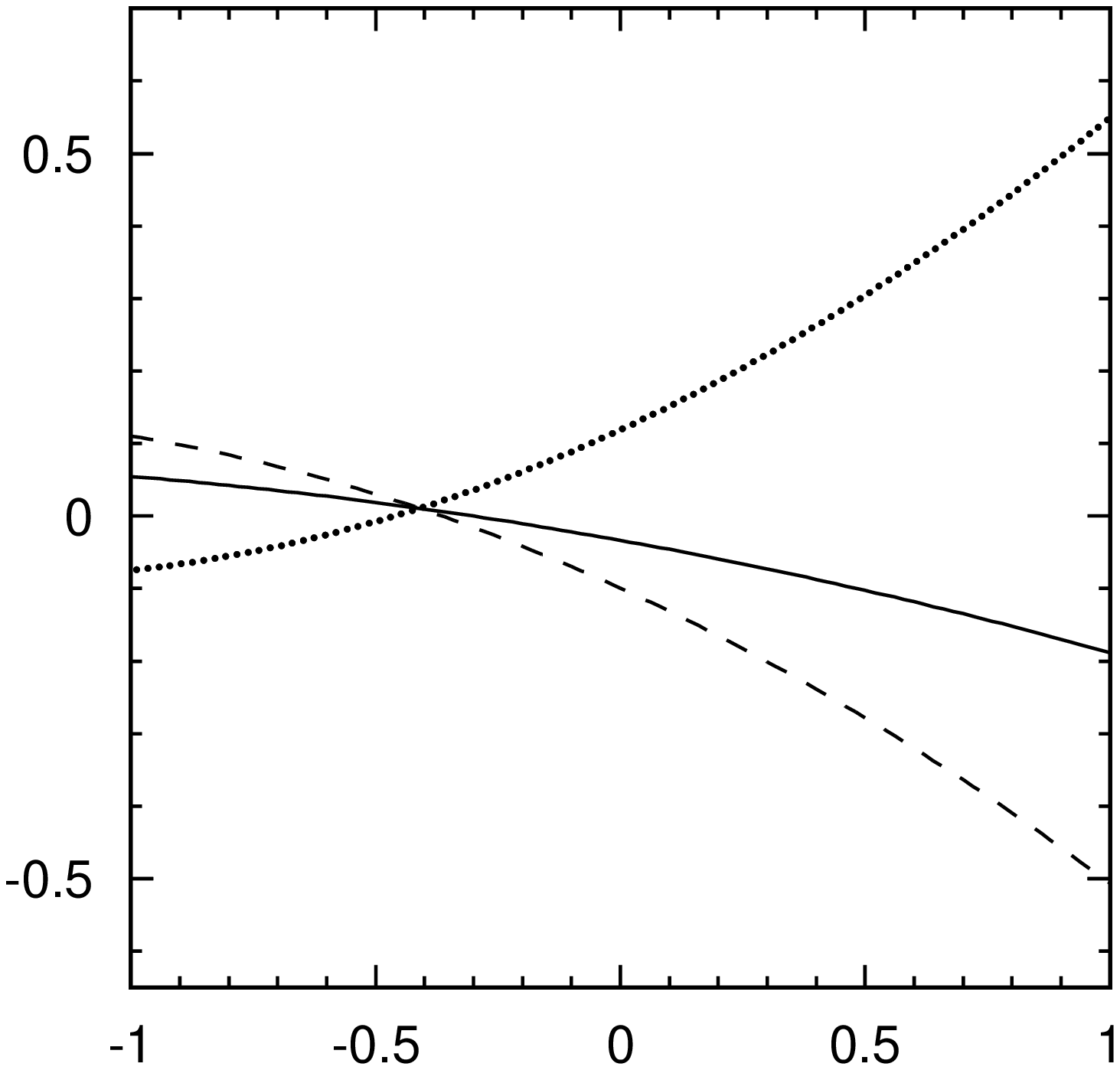,height=9cm,width=9cm}}
\put(8,-1){\psfig{figure=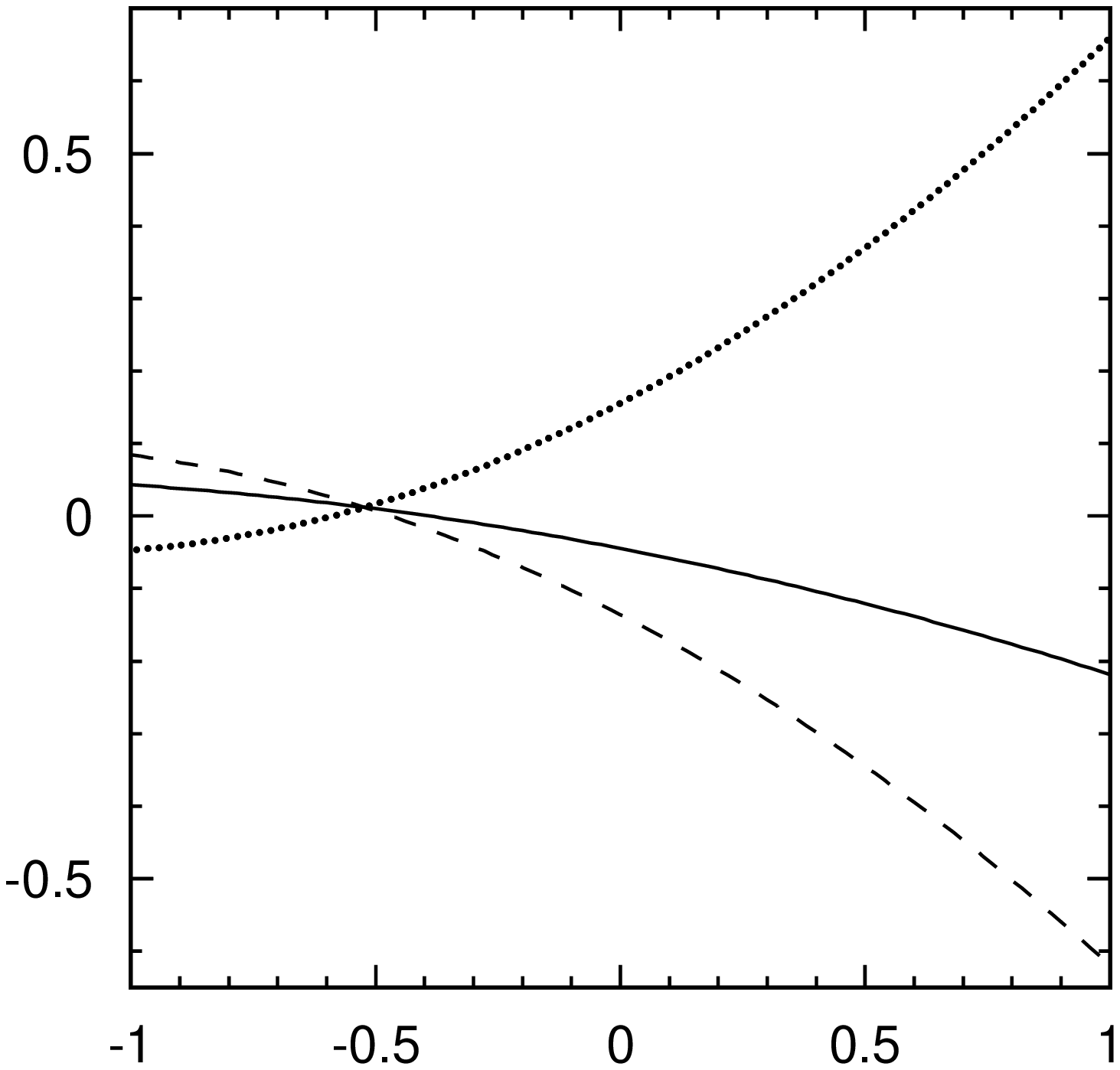,height=9cm,width=9cm}}
\put(0.6,6.2){\mbox{a)}}
\put(9.6,6.2){\mbox{b)}}
\put(3.4,-0.6){\mbox{$z$}}
\put(12.4,-0.6){\mbox{$z$}}
\end{picture}
\vskip 0.5cm
\caption{}
\end{figure}
\begin{figure}[h]
\unitlength1.0cm
\begin{picture}(8.,8.)
\put(-1,-1){\psfig{figure=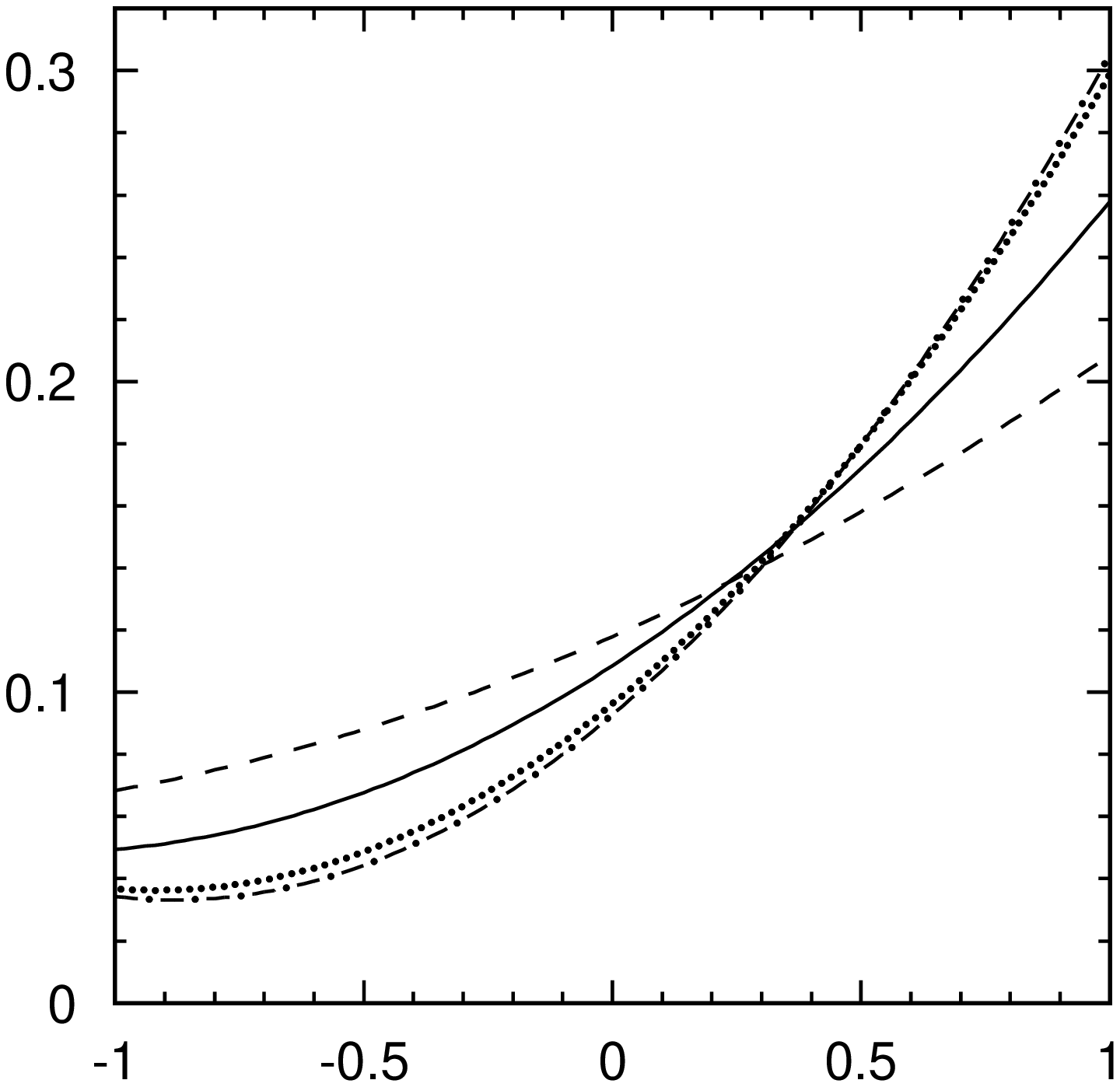,height=9cm,width=9cm}}
\put(3.4,-0.6){\mbox{$z$}}
\end{picture}
\vskip 0.5cm
\caption{}\label{fig:o4}
\end{figure}
\begin{figure}[h]
\unitlength1.0cm
\begin{picture}(8.,8.)
\put(-1,-1){\psfig{figure=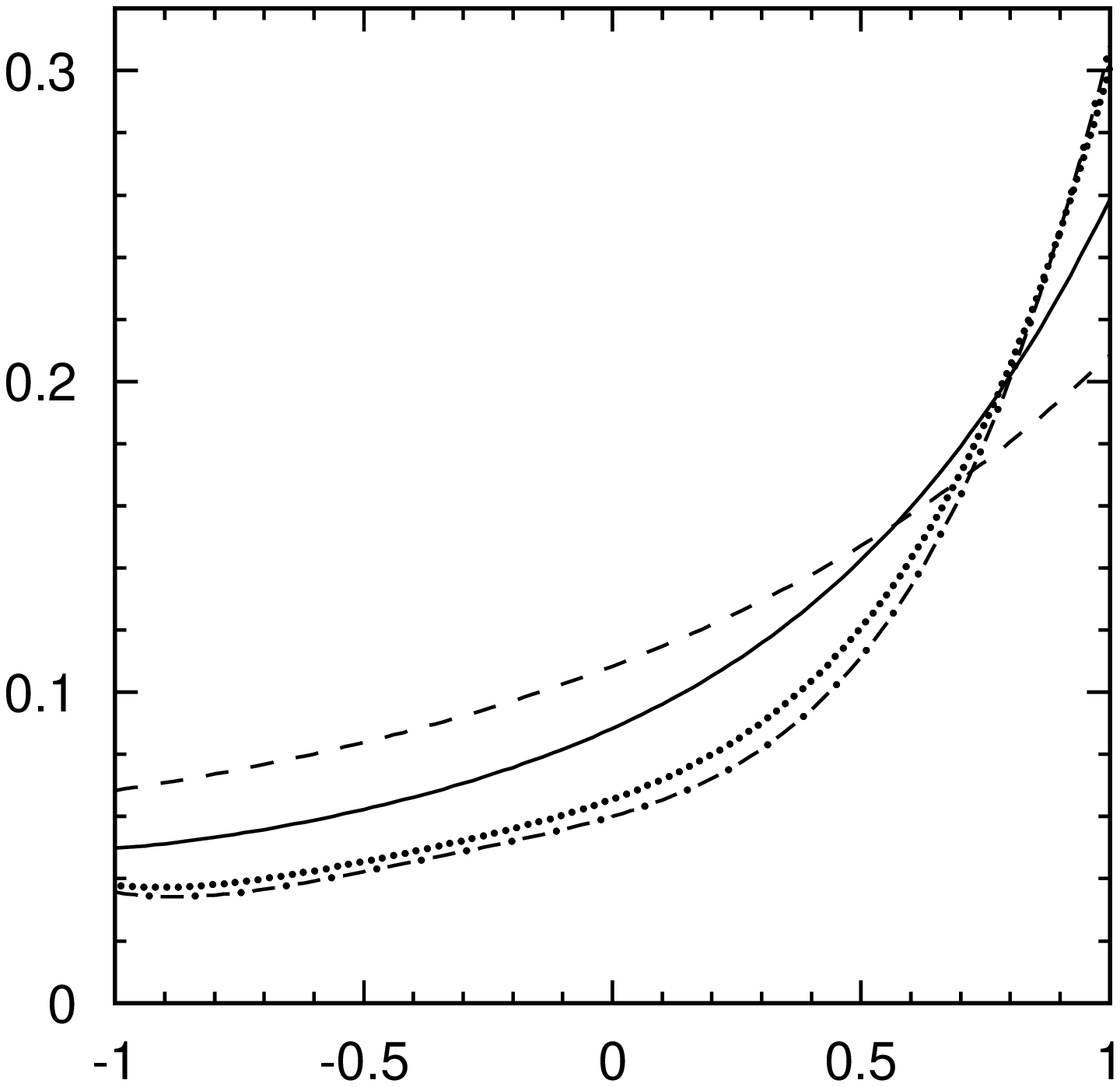,height=9cm,width=9cm}}
\put(3.4,-0.6){\mbox{$z$}}
\end{picture}
\vskip 0.5cm
\caption{}
\end{figure}
\begin{figure}[h]
\unitlength1.0cm
\begin{picture}(8.,8.)
\put(-1,-1){\psfig{figure=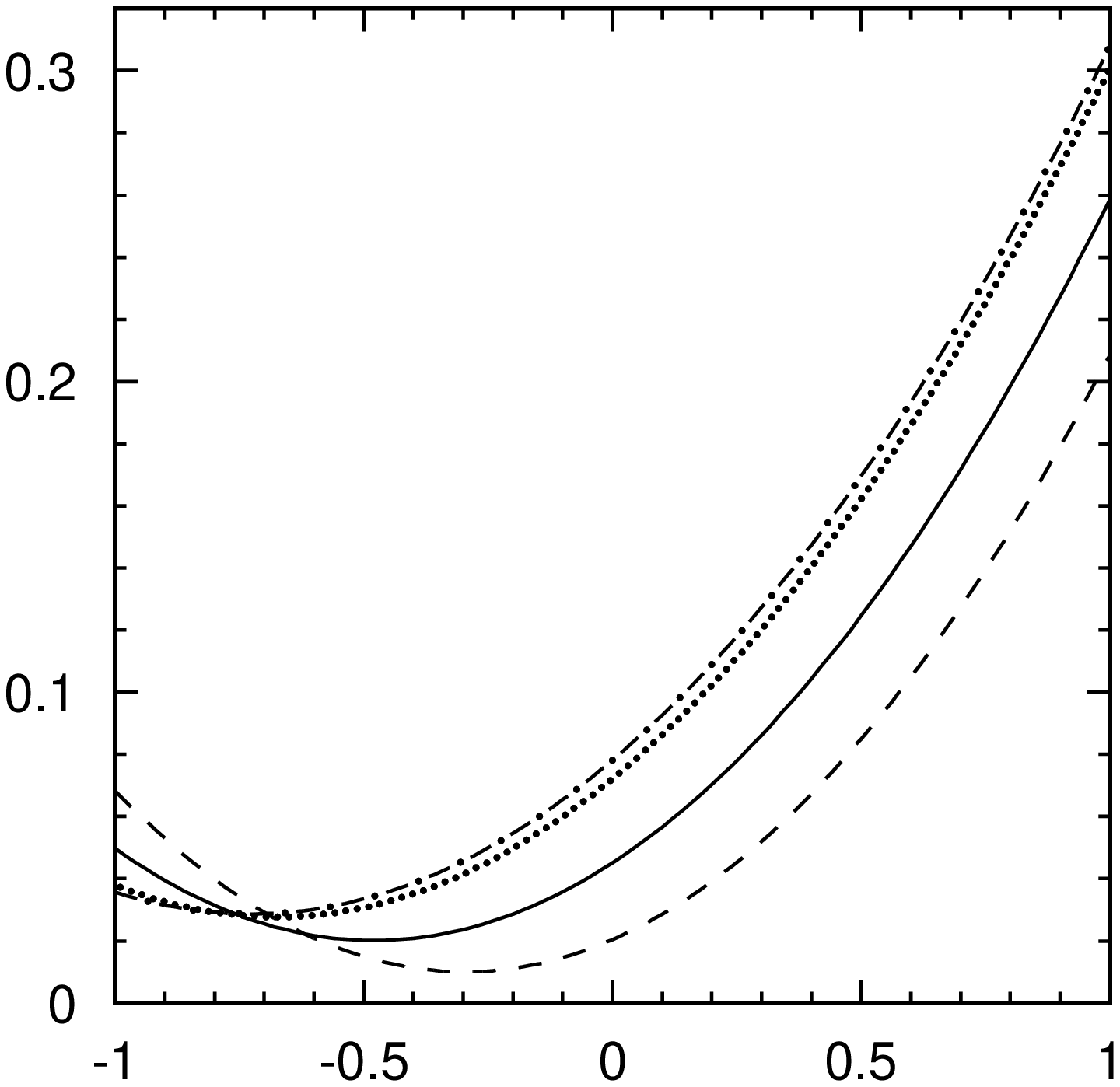,height=9cm,width=9cm}}
\put(3.4,-0.6){\mbox{$z$}}
\end{picture}
\vskip 0.5cm
\caption{}
\end{figure}
\begin{figure}[h]
\unitlength1.0cm
\begin{picture}(8.,8.)
\put(-1,-1){\psfig{figure=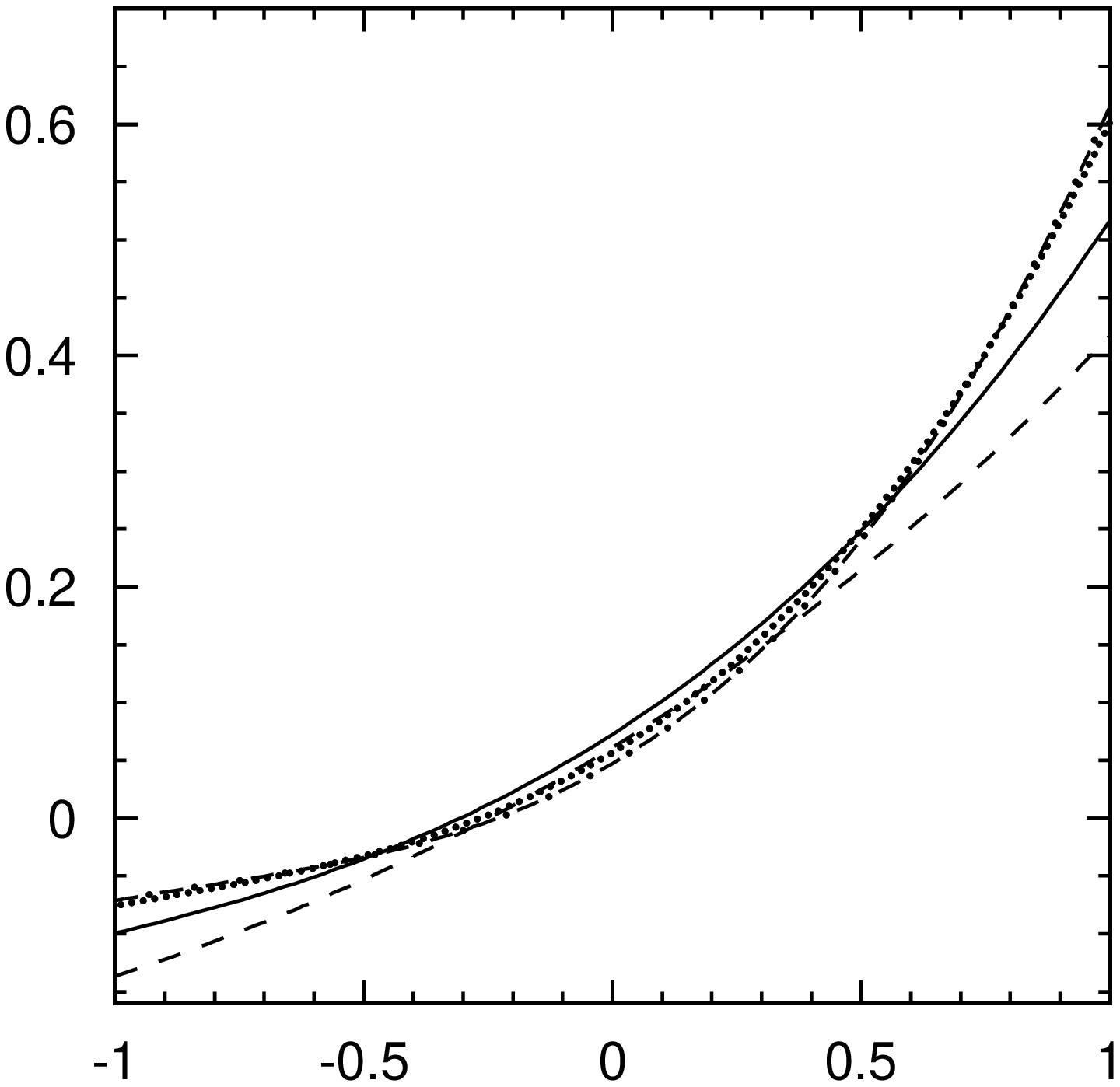,height=9cm,width=9cm}}
\put(3.4,-0.6){\mbox{$z$}}
\end{picture}
\vskip 0.5cm
\caption{}
\end{figure}
\begin{figure}[h]
\unitlength1.0cm
\begin{picture}(16.,8.)
\put(-1,-1){\psfig{figure=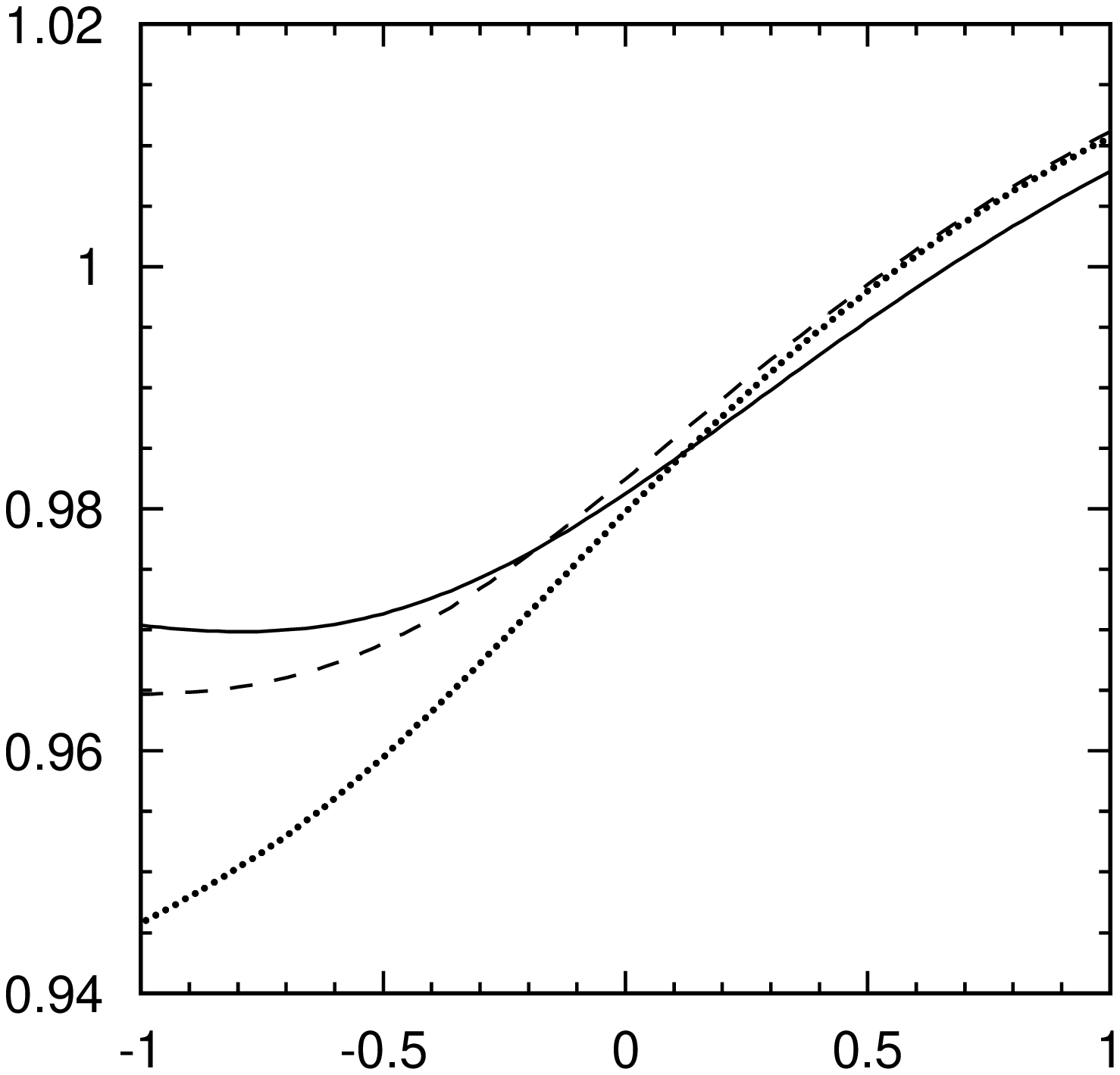,height=9cm,width=9cm}}
\put(8,-1){\psfig{figure=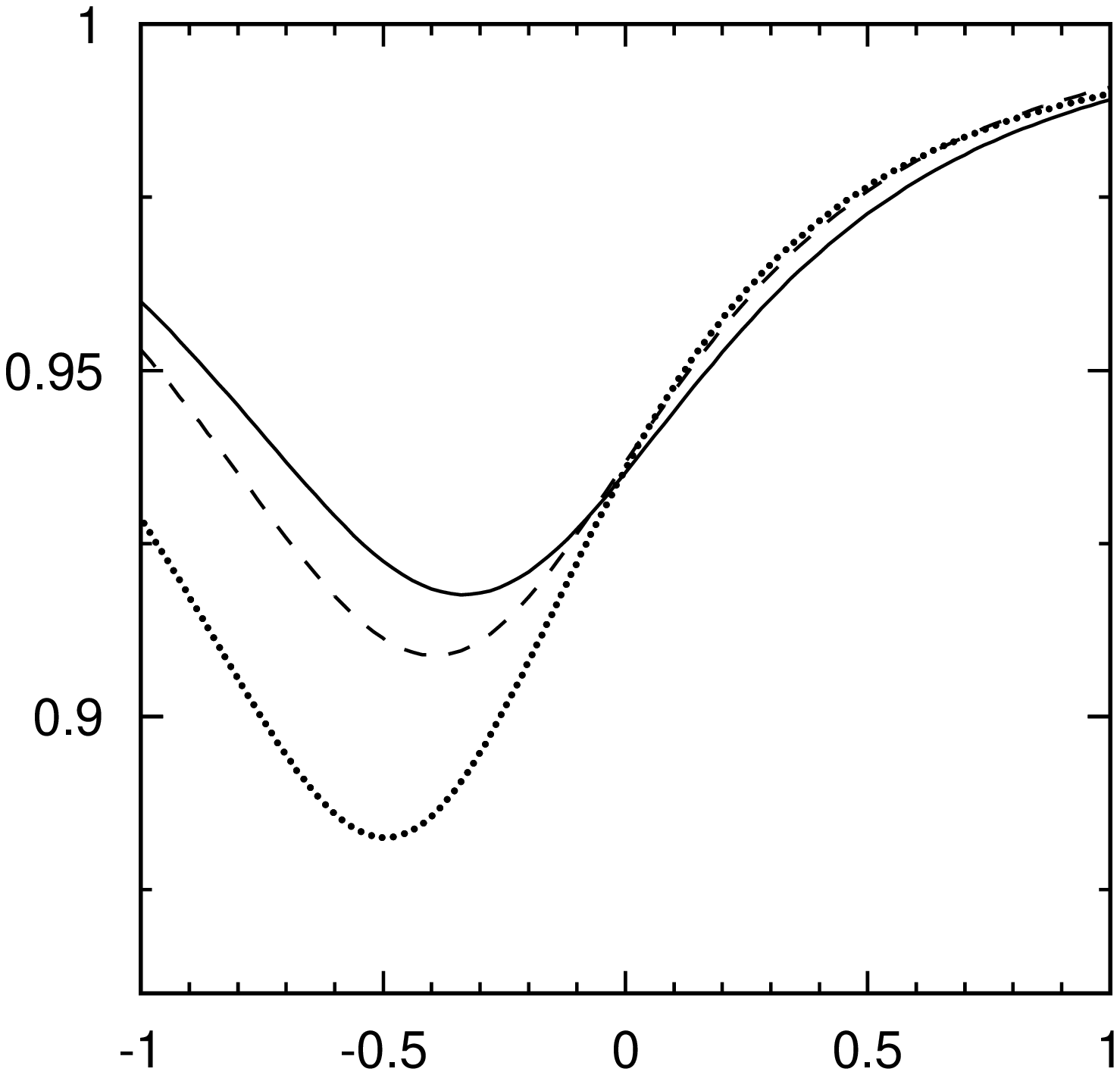,height=9cm,width=9cm}}
\put(0.6,6.2){\mbox{a)}}
\put(9.6,6.2){\mbox{b)}}
\put(3.4,-0.6){\mbox{$z$}}
\put(12.4,-0.6){\mbox{$z$}}
\end{picture}
\vskip 0.5cm
\caption{}
\end{figure}
\begin{figure}[h]
\unitlength1.0cm
\begin{picture}(8.,8.)
\put(-1,-1){\psfig{figure=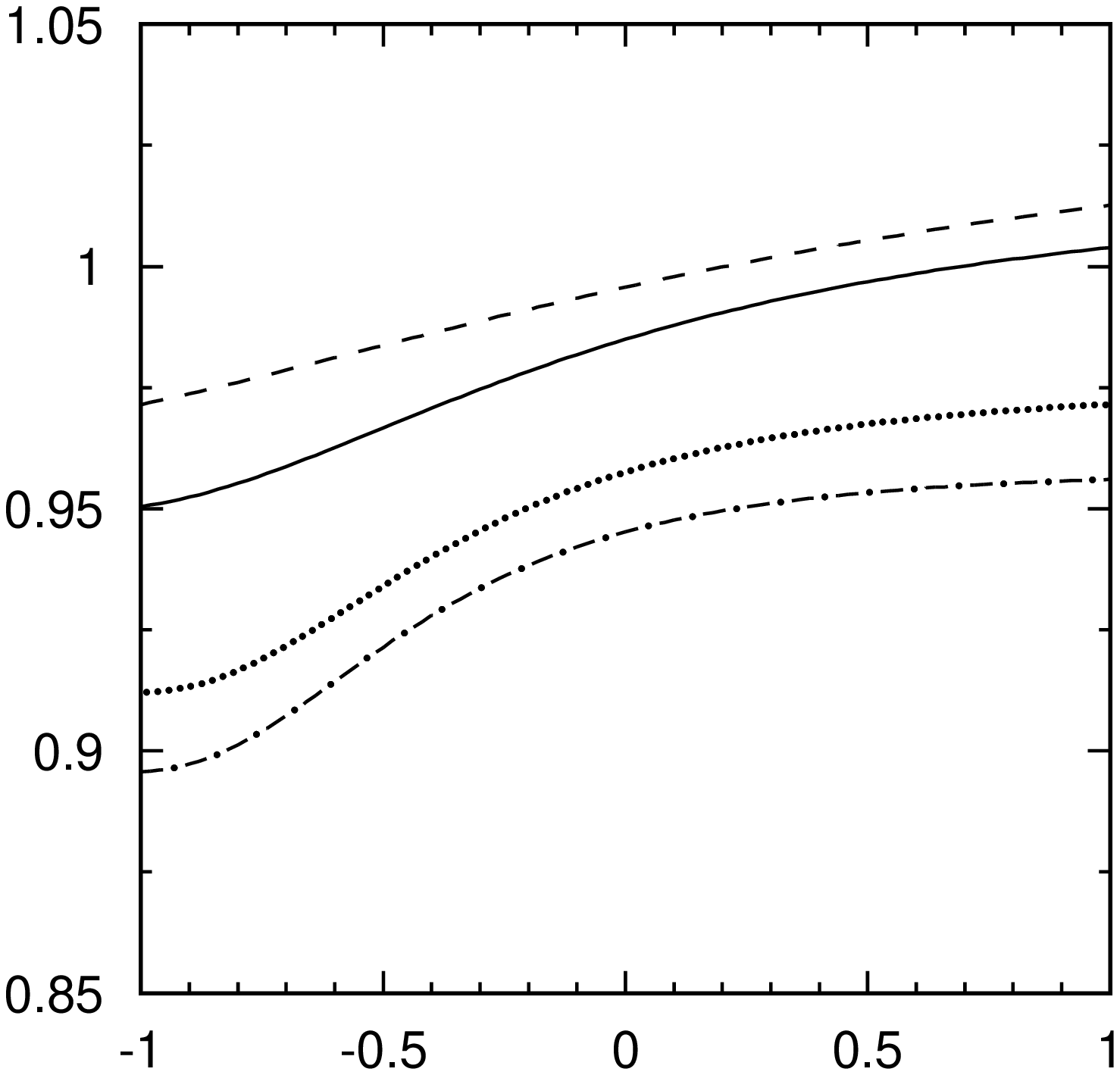,height=9cm,width=9cm}}
\put(3.4,-0.6){\mbox{$z$}}
\end{picture}
\vskip 0.5cm
\caption{}
\end{figure}
\begin{figure}[h]
\unitlength1.0cm
\begin{picture}(8.,8.)
\put(-1,-1){\psfig{figure=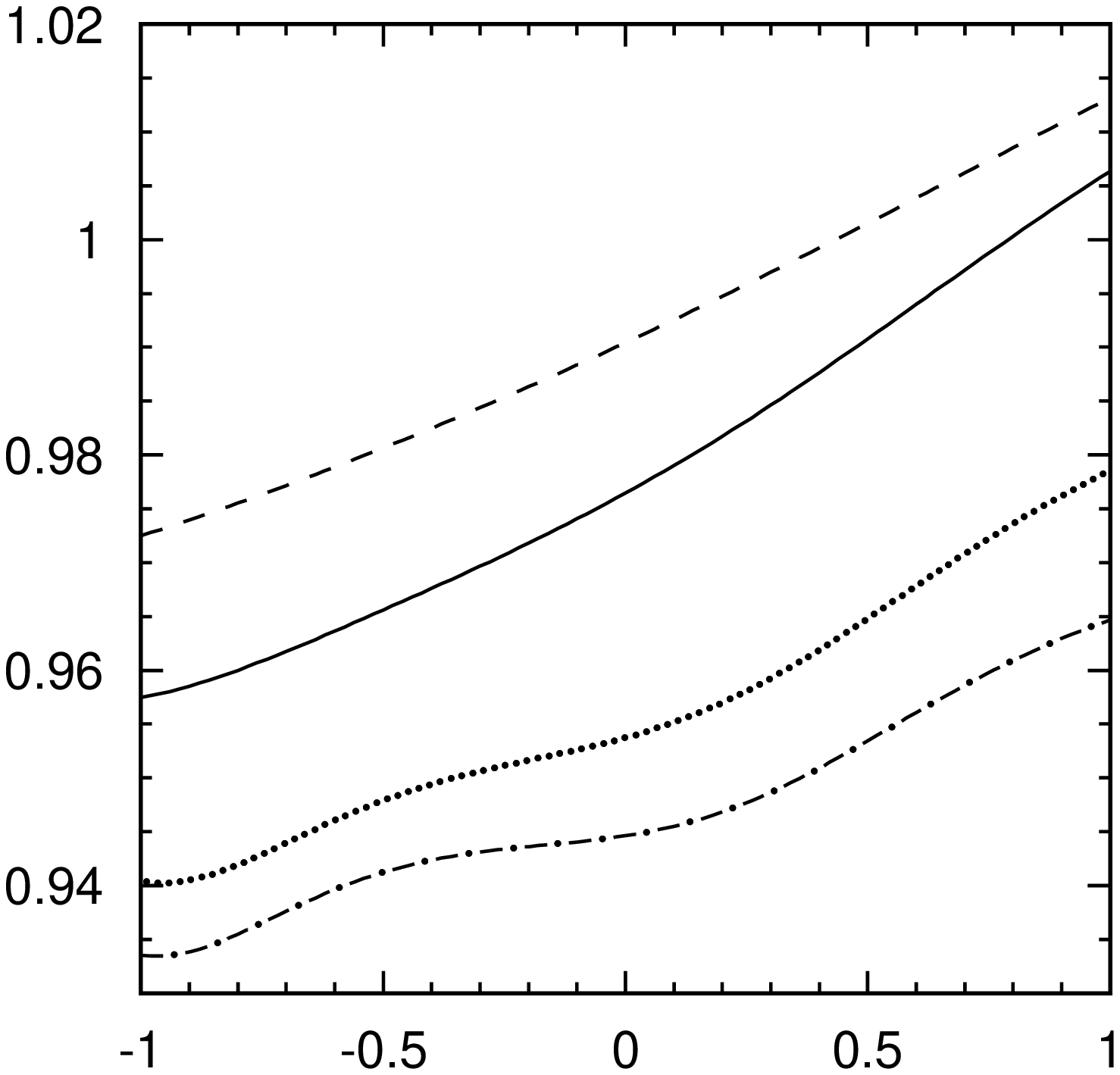,height=9cm,width=9cm}}
\put(3.4,-0.6){\mbox{$z$}}
\end{picture}
\vskip 0.5cm
\caption{}
\end{figure}
\begin{figure}[h]
\unitlength1.0cm
\begin{picture}(8.,8.)
\put(-1,-1){\psfig{figure=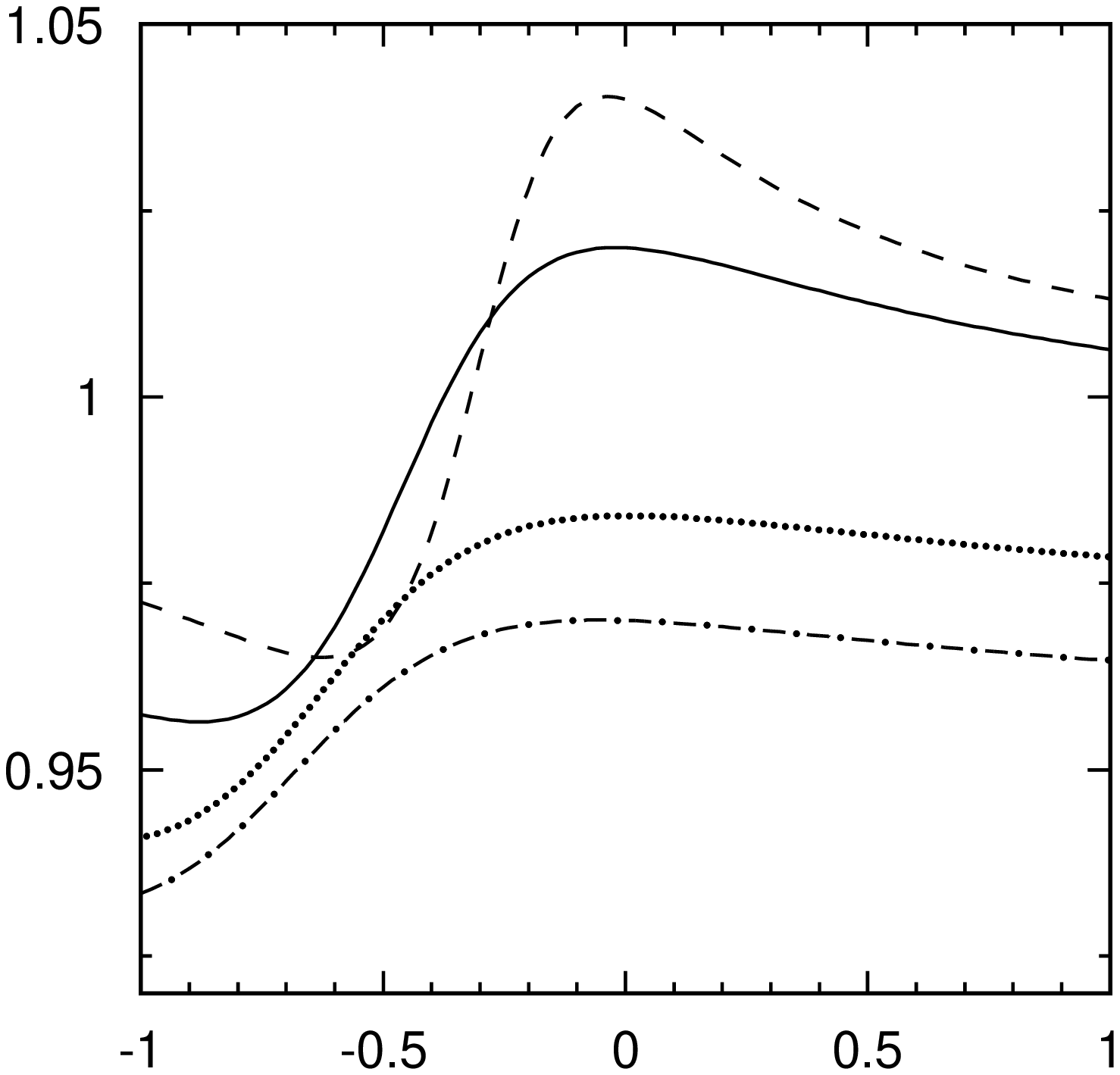,height=9cm,width=9cm}}
\put(3.4,-0.6){\mbox{$z$}}
\end{picture}
\vskip 0.5cm\caption{}
\end{figure}
\end{center}
\end{document}